\documentclass[aps,pre,reprint]{revtex4-1} 
\usepackage{amsfonts}
\usepackage{amsmath}
\usepackage{amssymb}
\usepackage{graphicx}
\usepackage{mathrsfs} 
\usepackage{color} 

\providecommand\bnabla{\boldsymbol{\nabla}}

\providecommand\bcdot{\boldsymbol{\cdot}}

\providecommand\bx{\mathbf{x}}

\providecommand\unit{\boldsymbol{\hat{\imath}}}

\newcommand{\pd}[2]{\frac{\partial #1}{\partial #2}}

\newcommand{\ub}[1]{^{({#1})}}

\begin{document} 
\title{Singular perturbations approach to localized surface-plasmon resonance: Nearly touching metal nanospheres}
\author{Ory Schnitzer}
\affiliation{Department of Mathematics, Imperial College London, South Kensington Campus, London SW7 2AZ, United Kingdom}

\begin{abstract}
Metallic nano-structures characterised by multiple geometric length scales support low-frequency surface-plasmon modes, which enable strong light localization and field enhancement. We suggest studying such configurations using singular perturbation methods, and demonstrate the efficacy of this approach by considering, in the quasi-static limit, a pair of nearly touching metallic nano-spheres subjected to an incident electromagnetic wave polarized with the electric field along the line of sphere centers. Rather than attempting an exact analytical solution, we construct the pertinent (longitudinal) eigen-modes by matching relatively simple asymptotic expansions valid in overlapping spatial domains. We thereby arrive at an effective boundary eigenvalue problem in a half-space representing the metal region in the vicinity of the gap. Coupling with the gap field gives rise to a mixed-type boundary condition with varying coefficients, whereas coupling with the particle-scale field enters through an integral eigenvalue selection rule involving the electrostatic capacitance of the configuration. By solving the reduced problem we obtain accurate closed-form expressions for the resonance values of the metal dielectric function. 
Furthermore, together with an energy-like integral relation, the latter eigen-solutions yield also closed-form approximations for the induced-dipole moment and gap-field enhancement under resonance. We demonstrate agreement between the asymptotic formulas and a semi-numerical computation. The analysis, underpinned by asymptotic scaling arguments, elucidates how metal polarization together with geometrical confinement enables a strong plasmon-frequency redshift and amplified near-field at resonance. \end{abstract}
\maketitle

\section{Introduction}
Advances in micro-fabrication technology along with new innovative ideas in nano-photonics and meta-materials have rekindled interest in the optics of small metal particles and structures \cite{Maier:05,Maier:07,Klimov:14}. Nano-metallic configurations exhibit unique optical characteristics, which are largely associated with the excitation of localized surface-plasmon (LSP) modes: conjoint resonant oscillations of electron-density and electromagnetic fields. LSP resonances are manifested in spectral peaks of the optical cross-sections and in significant near-field enhancements. Such phenomena are attractive for sensing applications \cite{Sonnichsen:05}, plasmonic manipulation of light below the diffraction limit \cite{Sukharev:07},  enhanced Raman scattering \cite{Xu:00,Kneipp:06}, plasmonic rulers \cite{Tabor:08}, lasers \cite{Oulton:09} and nanoantennas \cite{Muhlschlegel:05,Muskens:07b,Novotny:11}, second \cite{Hubert:07} and higher-order \cite{Kim:08} harmonic generation, and photovoltaics \cite{Atwater:10}. 

LSP resonance is exemplified by light scattering from a nano-metric metal sphere. In the quasi-static limit, the near-field, which appears to be driven by a uniform time-harmonic field, is derived from an electric potential governed by Laplace's equation. The induced electric-dipole moment is readily found to be  proportional to the Clausius--Mossotti factor $(\epsilon-1)/(\epsilon+2)$, where $\epsilon$ is the frequency-dependent dielectric function of the metal particle relative to its dielectric surrounding. The evident divergence at the Fr\"ohlich value $\epsilon=-2$  is associated with the existence of a solution to the homogeneous quasi-static problem (in which the incident field is absent). At visible and ultraviolet frequencies, a good approximation for the Drude dielectric function is \cite{Bohren:Book}
\begin{equation}\label{Drude}
\mathrm{Re}[\epsilon]\approx 1-\omega_p^2/\omega^2, \quad \mathrm{Im}[\epsilon]\approx \omega_p^2\gamma/\omega^3,
\end{equation}
where $\omega$ is the frequency of light, and $\omega_p$ and $\gamma$ are the plasma and collision frequencies, respectively (representative values for noble metals: $\hbar\omega_p\approx 9\,\mathrm{eV}$,  $\hbar\gamma\approx 0.13\, \mathrm{eV}$). Owing to the smallness of $\gamma/\omega_p$, a damped resonance occurs when $\mathrm{Re}[\epsilon]\approx-2$. 

LSP frequencies and the magnitude of resonant enhancement can be ``designed'' by shape and material variations \cite{Sosa:03,Myroshnychenko:08,Rueting:10,Chuntonov:11}. Appreciable modification is achievable with ``singular'' configurations characterised by non-smoothness or multiple-scale geometries:  sharp corners \cite{Dobrzynski:72,luo:12}; elongated particles \cite{Aizpurua:05,Andersen:02, Guzatov:11}; tips \cite{Wang:13}; nearly touching \cite{Aubry:10A,Aubry:11,luo:14}, touching \cite{Paley:93,Fernandez:10}, and embedded \cite{Jung:12} particle pairs; ``plasmonic molecules'' \cite{Chuntonov:11}; and particle-wall configurations \cite{Lei:12}. A recurring feature encountered in multiple-scale geometries is the existence of modes, that with increasing scale disparity, significantly redshift to low frequencies and large $-\mathrm{Re}[\epsilon]$ values. 

Singular configurations of the latter sort, which enable a particularly strong near-field enhancement \cite{Ciraci:12}, have been extensively studied through numerical simulations \cite{Ruppin:82,Mayergoyz:05,Romero:06,Giannini:07,Encina:10,Hohenester:12}, analytic solutions in separable coordinate systems \cite{Dobrzynski:72,Paley:93,Dutta:08}, and the powerful method of transformation optics (TO) \cite{Kraft:14,Pendry:15}. In two dimensions TO furnishes explicit solutions \cite{Aubry:10A,Aubry:11,luo:12} that can be generalised to address effects of retardation \cite{Aubry:10} and non-locality \cite{Fernandez:12}. TO is less straightforward to apply in three dimensions, and typically furnishes an efficient computational scheme \cite{Pendry:13} rather than simple analytic expressions; nevertheless, the method has been successfully applied in various scenarios, e.g.~in the design of broadband plasmonic structures \cite{Fernandez:10} and in the study of van der Waals forces \cite{luo:14}. Yet another approach, motivated by general scaling arguments \cite{Lebedev:10}, is based on approximate reductions of  implicit infinite-series solutions for nano-wire and nano-sphere dimers \cite{Klimov:07,Vorobev:10,Lebedev:13}; we delay a critical discussion of the accuracy of the expressions in Refs.~\onlinecite{Klimov:07,Lebedev:13} to \S\ref{sec:discussion}. 

The above methodologies are ``top-bottom'' --- necessarily relying on ``exact'' analytical or numerical solutions, which then need to be reduced or evaluated in singular geometric limits (e.g.~small gaps for dimers, increasing aspect ratio for elongated particles). In this paper, we suggest an alternative ``bottom-up'' approach, which is specifically suitable for analyzing the extreme physical characteristics of LSP resonance in multiple-scale metallic nano-structures. The approach is built upon the paradigm of singular perturbation methods, where an asymptotic rather than an exact solution is sought by carrying out an \textit{ab initio} singular analysis. In particular, applying asymptotic scaling arguments in conjunction with the method of matched asymptotic expansions \cite{Hinch:91} allows exploiting the spatial non-uniformity of the singular LSP modes towards systematically decomposing the problem governing these into several considerably simpler ones. The main outcomes are asymptotic expressions for the resonance frequencies, along with a minimalistic description of the corresponding  modes. In the context of excitation problems, when losses are low an asymptotic approximation for the field enhancement may be obtained with little additional effort.

To demonstrate this approach, we shall focus in this paper on the prototypical configuration of a pair of {nearly touching} metal nano-spheres \cite{Ruppin:82,Ruppin:89, Tamaru:02,Rechberger:03,Gunnarsson:05,Romero:06,Klimov:07,Jain:07,Encina:10,Lebedev:10,Lebedev:13,Kadkhodazadeh:14}. We limit ourselves to a quasi-static treatment of the near field, and focus on the axisymmetric and longitudinal (bonding) modes that singularly redshift with vanishing gap width, and the excitation of these by a long-wave electromagnetic plane wave polarized with the electric field along the line of centers. The paper proceeds as follows. In \S\ref{sec:formulation} we formulate the problem and note useful generalities. In \S\ref{sec:resonant} we consider the plasmonic eigenvalue problem and obtain closed-form expressions for the dielectric-function resonance values. In \S\ref{sec:forced} we employ the eigenvalues and eigen-potentials found in \S\ref{sec:resonant} to obtain formulas for the induced dipole moment and field enhancement in the gap under resonance conditions. In \S\ref{sec:num} we present a validation of our results against a semi-numerical solution of the quasi-static problem. Lastly, in \S\ref{sec:discussion} we recapitulate our results, compare with related works, and discuss recent developments along with future directions.

\begin{figure}[t]
   \centering
   \includegraphics[scale=0.45]{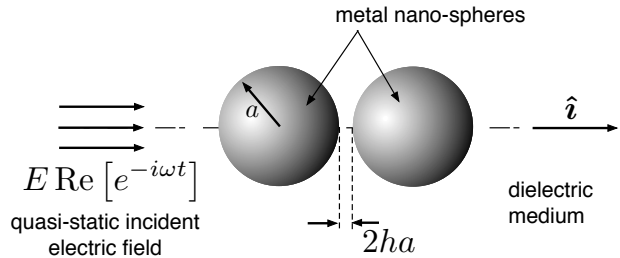} 
   \caption{Schematics of the quasi-static excitation problem.}
   \label{fig:dim_schematic}
\end{figure}

\section{Preliminaries} 
\label{sec:formulation}
\subsection{Problem formulation}
Consider an identical pair of metal spheres surrounded by a dielectric medium. The spheres are characterised by their radius $a$, and their frequency-dependent dielectric function relative to the background material, $\epsilon_0\epsilon$, $\epsilon_0$ being the vacuum permittivity. The minimum separation between the spheres is denoted by $2h a$. The pair is subjected to an incident time-harmonic electromagnetic plane wave of frequency $\omega$, polarized such that the electric field is in a direction $\unit$ parallel to the line of sphere centers, see Fig.~\ref{fig:dim_schematic}. We shall focus on the quasi-static limit of sufficiently small particles, $a\ll c/(\omega\sqrt{|\epsilon|})$,  $c$ being the speed of light in the surrounding dielectric \cite{Landau:Book,Bohren:Book}. In this description, the pair experiences the incident wave as a uniform time harmonic field $\unit E\mathrm{Re}[e^{-i\omega t}]$, where $E$ is a real constant magnitude, and the electric near-field is approximately curl-free, derived from a time-harmonic potential $aE\mathrm{Re}[\varphi e^{-i\omega t}]$. 

\begin{figure}[t]
   \centering
   \includegraphics[scale=0.43]{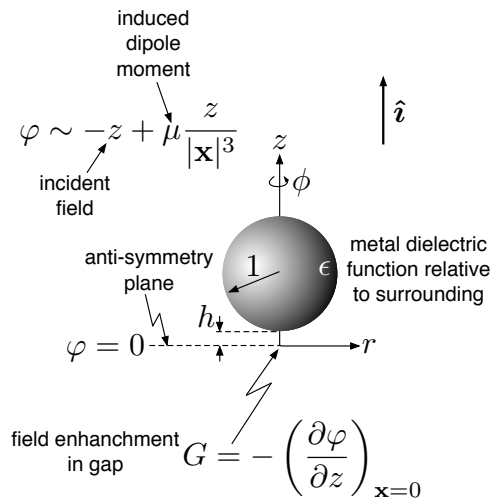} 
   \caption{Schematics of the dimensionless excitation problem.}
   \label{fig:nondim_schematic}
\end{figure}

Henceforth, we normalise all lengths by $a$. In particular, we denote by $\bx$ the dimensionless position vector relative to the centre of the gap. It is convenient to employ a cylindrical coordinate system $(r,z,\phi)$, with origin at $\bx=0$ and a $z$-axis directed along $\unit$ (see Fig.~\ref{fig:nondim_schematic}). The problem at hand is to find a continuous potential $\varphi$ that satisfies 
\begin{equation} \label{laplace gen}
\bnabla\bcdot\left(\mathcal{E}\bnabla\varphi\right)=0,
\end{equation}
where $\mathcal{E}$ is unity in the dielectric medium and $\epsilon$ in the metal spheres, 
together with the far-field condition 
\begin{equation} 
\varphi\sim- z+o(1) \quad \text{as} \quad |\bx|\to\infty. \label{far}
\end{equation}
The additive freedom of $\varphi$ is removed in \eqref{far} by choosing the disturbance from the incident field to attenuate at large distances. 

\subsection{General properties}
The following generalities can be inferred from \eqref{laplace gen}--\eqref{far}, see also Refs.~\cite{Mayergoyz:05,Klimov:14,Ando:14,Grieser:14,Ammari:15}. First, the potential is axisymmetric about the $\unit$ axis, i.e. $\partial\varphi/\partial\phi=0$. In addition, the potential is antisymmetric about the plane $z=0$. It is therefore possible to consider only the half-space $z>0$, with 
\begin{equation}
\varphi =0 \quad \text{at} \quad z=0. \label{as}
\end{equation} 
We next note that each of the spheres is overall charge free. That is, 
\begin{equation} \label{constraint}
\oint\pd{\varphi}{n}\,dA =0,
\end{equation}
where the integration is over an arbitrary closed surface entirely within the dielectric medium, which encloses one or both spheres. Thus $\varphi$ lacks a monopole term in its far-field expansion  \eqref{far}, which can therefore be extended to
\begin{equation} \label{far with mu}
\varphi\sim -z +\mu\frac{z}{|\bx|^3} \quad \text{as} \quad |\bx|\to\infty.
\end{equation}
The complex-valued constant $\mu$ is the induced dipole moment of the pair, normalised by $4\pi a^3\epsilon_0\epsilon_d E$ ($\epsilon_d$ being the relative dielectric constant of the surrounding medium). 


Green identities applied to $\varphi$ and its complex conjugate $\varphi^*$, in conjunction with \eqref{laplace gen} and \eqref{far with mu}, can be shown to give
\begin{equation}
4\pi \mathrm{Im}[\mu]=\mathrm{Im}[\epsilon] \int_{\text{metal}}\bnabla\varphi \bcdot \bnabla\varphi^*\,dV, \label{work}
\end{equation}
where the integration is over the volume of the two spheres. Eq.~\eqref{work} is a work-dissipation relation: The work done by the incident field on the effective induced dipole of the dimer is lost to Ohmic dissipation in the metal. This relation serves to remind us that we are in the quasi-static limit, where radiation losses are negligible. We note that $4\pi\mathrm{Im}[\mu]$ gives the quasi-static approximation for the absorption cross-section normalised by $\omega a^3/c$ \cite{Maier:07}.

Excitation problems of the above sort are ill posed for a  set of discrete, real and negative, $\epsilon$ values, for which the homogeneous variant of the problem obtained by omitting the incident field in \eqref{far} possesses non-trivial solutions. Since in practice $\mathrm{Im}[\epsilon]\ne0$, such solutions are often termed ``virtual'' LSP modes. Consider now an eigen-potential $\varphi_r$ corresponding to one such eigenvalue $\epsilon_r$. At large distances,
\begin{equation} \label{resonance mu}
\varphi_r\sim \mu_r\frac{z}{|\bx|^3} \quad \text{as} \quad |\bx|\to\infty,
\end{equation}
where $\mu_r$ is arbitrary. We note that the integral relation \eqref{work} does not apply to $\varphi_r$, since condition \eqref{far} is used to derive it. Say, however, that $\varphi$ is the solution to the actual quasi-static problem, with $\epsilon=\epsilon_r+\Delta\epsilon$. Applying the Green identities to the pair of potentials $(\varphi,\varphi_r)$, one can easily show that 
\begin{equation}
4\pi\mu_r=-\Delta\epsilon\int_{\text{metal}}\bnabla\varphi_r \bcdot \bnabla\varphi\,dV. \label{work mod}
\end{equation}
Whereas \eqref{work} bears a clear physical significant, we shall find in \S\ref{sec:forced} that \eqref{work mod} is more useful for deriving quantitative estimates for the  resonant field enhancement.

\subsection{Near-contact limit}
Henceforth we consider the near-contact limit where $h\to0$.
Specifically, our goal will be to derive closed-form asymptotic formulas for the eigenvalues $\epsilon\ub{n}$, and the values at resonance conditions (i.e.~when $\mathrm{Re}[\epsilon]=\epsilon\ub{n}$) of the induced-dipole moment $\mu$ and the field enhancement in the gap, $G=-\left(\partial{\varphi}/\partial{z}\right)_{\bx=0}$.

\section{LSP modes} \label{sec:resonant}
In this section we consider the boundary eigenvalue problem obtained by replacing the far field \eqref{far} with the homogeneous condition $\varphi\to0$ as $|\bx|\to\infty$. 
It is clear that the only relevant modes are those that are both longitudinal, i.e.~anti-symmetric about the plane $z=0$, and axisymmetric. 

\subsection{Scaling arguments}
\label{ssec:scaling}
How do the eigenvalues $\epsilon$ scale with $h$? In the vicinity of the gap the spherical boundaries can be approximated by two oppositely facing paraboloids. The separation therefore remains $O(h)$ for $O(h^{1/2})$ radial distances. We shall refer to the subdomain of the dielectric medium where $z=O(h)$ and $r=O(h^{1/2})$ as the gap domain, and assume without loss of generality that the eigen-potential $\varphi$ corresponding to $\epsilon$ is $O(1)$ there; since $\varphi$ is antisymmetric about the $z$ plane, the transverse gap field must be $O(1/h)$. Consider now the particle-scale dielectric ``outer'' domain, away from the gap, where $\varphi$ varies over distances $|\bx|=O(1)$ and is at most $O(1)$. It follows from \eqref{as} that $\varphi=O(h)$ when $z=O(h)$ and $r=O(1)$. This implies that the gap potential must vary in the radial direction over $O\left(h^{1/2}\right)$  distances. 
Consider next the metal sphere in $z>0$, in the vicinity of the gap boundary. Continuity of potential implies that $\varphi$ is $O(1)$ there, and varies rapidly over $O\left(h^{1/2}\right)$ distances along the boundary. On the latter length scale, the metallic volume appears transversely semi-infinite. Thus, the symmetry of Laplace's equation implies comparably rapid transverse variations. We shall refer to the metal subdomain where $r,z=O(h^{1/2})$ as the pole domain. 

The scaling of $\epsilon$ now follows by considering the continuity of electric displacement across the gap-pole interface: The product of $\epsilon$ and the $O(h^{-1/2})$ transverse field in the metal pole must be on the order of the $O(1/h)$ transverse field in the gap. We accordingly find
\begin{equation} \label{eps scaling}
\epsilon\sim -  h^{-1/2}\lambda(h)+O(1), \quad 0<\lambda=O(1),
\end{equation}
where, as we shall see, the pre-factor $\lambda$ itself depends weakly upon $h$, being a function of $\ln(1/h)$. This more subtle feature can also be anticipated from scaling arguments, as we now discuss. 

The largeness of $|\epsilon|$ suggests that, on the particle-scale, the spheres are approximately equipotential. Thus, the particle-scale potential distribution in the dielectric medium is approximately that about a pair of nearly touching spheres held at uniform potentials, say $\pm \mathcal{V}$. But the latter distribution implies a net charge on each sphere, in contradiction with \eqref{constraint}. This means that the deviation in the vicinity of the gap of $\varphi$  from the particle-scale distribution must be consistent with a localized accumulation of charge such that the overall charge of each sphere is zero. In fact, the contributions to the integral in \eqref{constraint} from the gap and outer dielectric domains are readily seen to be comparable: The former is given by the integral of an $O(1/h)$ transverse field over an $O(h)$ surface area, while the latter is obviously $O(1)$. This implies a connection between the solution in the  gap and pole ``inner'' domains, and the capacitance of a pair of perfectly conducting spheres held at fixed potentials $\pm\mathcal{V}$, which is known from electrostatic theory to depend logarithmically upon $h$ as $h\to0$ \cite{Jeffrey:78}.

\begin{figure}[t]
   \centering
   \includegraphics[width=\columnwidth]{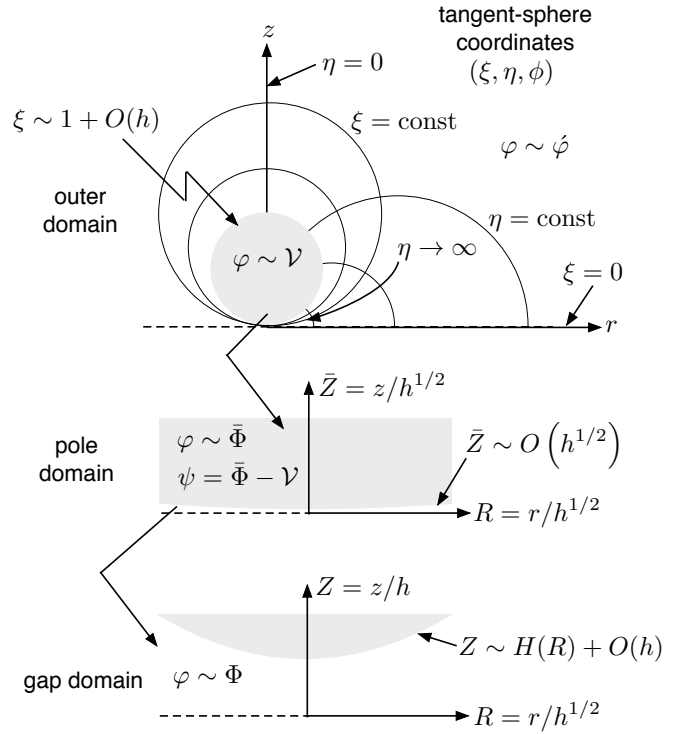} 
   \caption{The resonant modes are constructed in \S\ref{sec:resonant} by matching simple asymptotic expansions describing the gap, pole, and particle-scale ``outer'' domains.}
   \label{fig:domains}
\end{figure}
\subsection{Gap and pole domains}
We begin constructing the eigen-potentials by considering the gap domain. To this end, we introduce the stretched coordinates 
\begin{equation}
R=r/h^{1/2}, \quad Z=z/h.  \label{gap RZ}
\end{equation}
The gap domain is where $R,Z\sim O(1)$, with $R>0$ and $Z$ bounded by the spherical surfaces (see Fig.~\ref{fig:domains})
\begin{equation}
Z\sim \pm H(R) + O(h), \quad \text{where} \quad H(R)=1+\frac{1}{2}R^2.
\end{equation}
As in \S\ref{ssec:scaling}, we assume without loss of generality that $\varphi=O(1)$ in the gap domain. We accordingly pose the asymptotic expansion
\begin{equation}
\varphi\sim \Phi(R,Z) + O\left(h^{1/2}\right). \label{Phi exp}
\end{equation}
Substituting \eqref{Phi exp} into Laplace's equation, written in terms of $R$ and $Z$, yields at leading order
\begin{equation}
\pd{^2\Phi}{Z^2}=0. \label{Lap gap}
\end{equation}
Integration of \eqref{Lap gap} in conjunction with \eqref{as} gives
\begin{equation} \label{Phi}
\Phi=A(R)Z,
\end{equation}
where the function $A(R)$ remains to be determined. 

Consider next the pole domain of the metal sphere in $z>0$, described by the alternative pair of stretched coordinates 
\begin{equation}
R=r/h, \quad \bar{Z}=z/h^{1/2}. \label{pole RZ}
\end{equation}
The pole domain is where $R,\bar{Z}\sim O(1)$, with $R>0$ and $\bar{Z}$ bounded from below by the sphere surface $\bar{Z}\sim O(h^{1/2})$; under the rescaling \eqref{pole RZ}, the pole domain is to leading order a semi-infinite half space. We expand the pole-domain potential as 
\begin{equation}
\varphi\sim\bar{\Phi}(R,\bar{Z})+O\left(h^{1/2}\right), \label{barPhi}
\end{equation} 
where $\bar\Phi$ is governed by Laplace's equation,
\begin{equation} \label{pole laplace}
\frac{1}{R}\pd{}{R}\left(R\pd{\bar\Phi}{R}\right)+\pd{^2\bar\Phi}{\bar Z^2}=0.
\end{equation}
Expanding the eigenvalue $\epsilon$ according to \eqref{eps scaling}, requiring leading-order continuity of potential and electric displacement yields
\begin{equation} 
\Phi=\bar{\Phi}, \quad \pd{\Phi}{Z}+\lambda \pd{\bar\Phi}{\bar{Z}}=0 \quad \text{at} \quad Z=H(R),\,\bar{Z}=0. \label{pole conditions}
\end{equation}
Eqs.~\eqref{pole conditions} may be combined to yield the mixed-type boundary condition
\begin{equation} \label{mixed}
\bar{\Phi}+\lambda H \pd{\bar\Phi}{\bar{Z}}=0 \quad \text{at} \quad \bar{Z}=0,
\end{equation}
which involves $\bar\Phi$ alone. We also find the relation
\begin{equation}
A(R)=\frac{1}{H}\left(\bar{\Phi}\right)_{\bar{Z}=0}, \label{A}
\end{equation}
which provides the gap potential once the pole potential has been determined. As discussed in \S\ref{ssec:scaling}, the particle-scale metal domain  is equipotential to leading order. Asymptotic matching thus furnishes the ``far-field'' condition
\begin{equation} \label{pole far}
\bar\Phi\to \mathcal{V} \quad \text{as} \quad R^2+\bar{Z}^2\to\infty,
\end{equation}
where $\mathcal{V}$ is the above-mentioned uniform potential of the sphere in $z>0$. For later reference, we note that \eqref{A} and \eqref{pole far} imply $\Phi\sim 2\mathcal{V}Z/R^2$ as $R\to\infty$. 

The pole potential $\bar\Phi$ is governed by Laplace's equation \eqref{pole laplace} in the half space $\bar{Z}>0$, the boundary condition \eqref{mixed}, and the far-field condition \eqref{pole far}. We shall see in \S\ref{ssec:algebra} that this boundary-value problem possesses non-trivial solutions for essentially arbitrary $\lambda>0$ and $\mathcal{V}$, excluding a discrete set of $\lambda$ values for which a solution exists only if $\mathcal{V}$ is set to zero. It is tempting to guess that the latter set furnishes the requisite eigenvalues; we shall see in \S\ref{ssec:logarithmic} however that these only give a leading term in an expansion of $\lambda$ in inverse powers of $\ln(1/h)$. Avoiding such gross logarithmic errors, the eigenvalues cannot be solely determined from local considerations, as already anticipated in \S\ref{ssec:scaling}. We accordingly divert at this stage to consider the particle-scale dielectric domain.

\subsection{Outer domains and global constraint}
In the outer limit, $h\to0$ with $\bx=O(1)$, the spheres appear to be in contact to leading order. It is therefore convenient to introduce the tangent-sphere coordinates defined through \cite{Moon:61}
\begin{equation} \label{tangent}
z=\frac{2\xi}{\xi^2+\eta^2}, \quad r=\frac{2\eta}{\xi^2+\eta^2}.
\end{equation}
The plane $z=0$ is mapped to $\xi=0$, the upper-half-space sphere to $\xi\sim 1 + O(h)$, and the origin to $\eta\to\infty$ (see Fig.~\ref{fig:domains}). We expand the potential in the outer dielectric domain as 
\begin{equation}
\varphi\sim\acute\varphi+O(h^{1/2}),
\end{equation} 
and in the outer metal domain as
\begin{equation}
\varphi\sim\pm\mathcal{V}+O(h^{1/2}),
\end{equation} 
the $\pm$ sign corresponding to the sphere in $z\gtrless0$, respectively. 

For $\acute\varphi$ we require a solution to Laplace's equation that attenuates at large distances and satisfies $\acute\varphi=\mathcal{V}$ at $\xi=1$ and $\acute\varphi=0$ at $\xi=0$. In addition, the solution must asymptotically match the gap and pole potentials. Standard Hankel-Transform methods yield the requisite potential as \cite{Jeffrey:78}
\begin{equation}
\acute\varphi\sim \mathcal{V}(\xi^2+\eta^2)^{1/2}\int_0^{\infty}e^{-s}\frac{\sinh(s\xi)}{\sinh s} J_0(s\eta)\,ds, \label{outer}
\end{equation}
where $J_0$ is the zeroth-order Bessel function of the first kind. It may be readily shown that $\acute\varphi\sim 2\mathcal{V}z/r^2$ as $\eta\to\infty$, whereby \eqref{outer} matches with the gap and pole domains. Asymptotic evaluation of \eqref{outer} for small $\xi$ and $\eta$, together with \eqref{tangent}, yields the asymptotic behaviour of \eqref{outer} at large distances [cf.~\eqref{resonance mu}],
\begin{equation} \label{muV}
\acute\varphi\sim \frac{\pi^2\mathcal{V}}{3}\frac{z}{|\bx|^3} \quad \text{as} \quad |\bx|\to\infty.
\end{equation}

The leading-order matching of \eqref{outer} with the solution in the gap and pole regions has been achieved, apparently without introducing any additional  constraint on the problem governing $\bar\Phi$. Instead of continuing the asymptotic analysis to the next order in search of the awaited eigenvalue selection rule, it proves simpler to follow the discussion in \S\ref{ssec:scaling} and consider the charge constraint \eqref{constraint} applied to the sphere in $z>0$. In accordance with the discussion there, an attempt to calculate the contributions to the integral in \eqref{constraint} from the gap and outer region separately results in diverging integrals --- this is typical in cases where contributions from overlapping asymptotic domains are comparable \cite{Hinch:91}. Following the analysis in Ref.~\cite{Jeffrey:78}, we write \eqref{constraint} as
\begin{equation} \label{constraint leading}
\int_0^{R_0}A(R)R\,dR + \int_0^{\eta_0}\left.\pd{\acute\varphi}{\xi}\right|_{\xi=1}\frac{2\eta}{1+\eta^2}\,d\eta = O\left(h^{1/2}\right),
\end{equation}
where $\eta_0,R_0\gg1$ are connected through $h^{1/2}R_0=2/\eta_0+O(1/\eta_0^3)$. The first integral in \eqref{constraint leading}, which represents the \textit{excess} contribution from the gap region, can be written as
\begin{equation}
\sim  \int_0^{\infty}\left[A(R)-\frac{\mathcal{V}}{H(R)}\right]R\,dR + \mathcal{V}\ln H(R_0).
\end{equation}
The second integral in \eqref{constraint leading}, which represents the contribution of the outer domain, can be worked out to be 
\begin{equation}
\sim \mathcal{V}\left[2\gamma_{e}+\ln(1+\eta_0^2)\right],
\end{equation}
where $\gamma_e\approx0.5772$ is the Euler--Gamma constant. Noting that $\ln H(R_0)\sim \ln({2}/{h})-\ln \eta_0^2$, and invoking \eqref{A}, the singular parts of the two contributions cancel out, and we find the anticipated ``global'' constraint:
\begin{equation} \label{V eq}
\mathcal{V}=-\frac{1}{\ln(2/h)+2\gamma_e}\int_0^{\infty}\frac{\bar\Phi(R,0)-\mathcal{V}}{H(R)}R\,dR.
\end{equation}

\subsection{Reduced eigenvalue problem}
\label{ssec:reduced}
Together with \eqref{V eq}, the problem governing the pole potential $\bar\Phi$ constitutes a reduced boundary-eigenvalue problem. It is convenient to recapitulate this problem in terms of the disturbance potential, $\psi(R,\bar{Z})=\bar\Phi-\mathcal{V}$. We have Laplace's equation
\begin{equation}
\frac{1}{R}\pd{}{R}\left(R\pd{\psi}{R}\right)+\pd{^2\psi}{\bar{Z}^2}=0 \label{reduced lap}
\end{equation}
in the upper half-space $\bar{Z}>0$, the integro-differential varying-coefficient boundary condition 
\begin{equation}\label{reduced bc}
{\psi}+\lambda H(R) \pd{\psi}{\bar{Z}}=\frac{1}{\ln(2/h)+2\gamma_e}\int_0^{\infty}\frac{\psi(R,0)}{H(R)}R\,dR
\end{equation}
at $\bar{Z}=0$ [we will sometimes write $-\mathcal{V}$ for the right-hand-side, see \eqref{V eq}], and the far-field condition
\begin{equation}\label{reduced far}
\psi\to0 \quad \text{as} \quad R^2+\bar{Z}^2\to\infty.
\end{equation}

The logarithmic dependence of the eigenvalues $\lambda$ and their corresponding eigen-potentials $\psi$ upon $h$ is evident from \eqref{reduced bc}. We also see that a rough leading-order solution, entailing an $O(\ln^{-1}\frac{1}{h})$ relative error, may be obtained by neglecting the right hand side of \eqref{reduced bc}. In the following subsections we proceed in two paths: (i) in \S\ref{ssec:logarithmic} we calculate succesive  terms in inverse logarithmic powers; (ii) in \S\ref{ssec:algebra} we tackle the reduced eigenvalue problem as a whole, thereby finding an ``algebraically accurate'' approximation for the resonance $\epsilon$ values. 

\subsection{Two terms in a logarithmic expansion}
\label{ssec:logarithmic}
We here expand the potentials $\psi$ and the eigenvalues $\lambda$ as
\begin{equation} \label{log exp}
\psi\sim \psi_0+\frac{1}{\ln\frac{1}{h}} \psi_1 + \cdots, \quad \lambda \sim \lambda_0 + \frac{1}{\ln\frac{1}{h}} \lambda_1+\cdots
\end{equation}
At leading order, the boundary condition \eqref{reduced bc} reads
\begin{equation} \label{bc 0}
\psi_0+\lambda_0 H({R})\pd{\psi_0}{\bar{Z}}=0 \quad \text{at} \quad \bar{Z}=0.
\end{equation}
We look for a solution in the form
\begin{equation} \label{hank 0}
\hat{\psi_0}(s,\bar Z)=\mathcal{H}\left[\psi_0(R,\bar Z)\right],
\end{equation}
where 
\begin{equation} \label{hank def}
\mathcal{H}[f(R)](s)= \int_0^{\infty}f(R)J_0(sR)R\,dR
\end{equation}
is the zeroth-order Hankel transform \cite{Sneddon:Book}. Applying the Hankel transform to Laplace's equation \eqref{reduced lap}, a solution that satisfies \eqref{reduced far} is
\begin{equation} \label{psi 0}
\hat{\psi_0}=\frac{1}{s}{Y_0(s)}e^{-s\bar{Z}},
\end{equation}
where $Y_0(s)$ depends on the conditions at $\bar{Z}=0$. Transforming the boundary condition \eqref{bc 0}, we find the $s$-space differential equation
\begin{equation} \label{log Y}
\frac{1}{s}\frac{d}{ds}\left(s\frac{dY_0}{ds}\right)+2\left(\frac{1}{\lambda_0 s }-1\right)Y_0=0.
\end{equation}
The first term in \eqref{log Y} is obtained through integration by parts twice, which is only applicable if \cite{Sneddon:Book}
\begin{equation} \label{IBP conditions}
Y_0\ll \frac{1}{s^{2}} \quad \text{and} \quad \frac{dY_0}{ds}\ll \frac{1}{s} \quad  \text{as} \quad s\to0.
\end{equation}
In addition, $Y_0(s)$ must decay sufficiently fast to allow convergence of \eqref{hank 0}. 
Defining
\begin{equation} \label{Y to T}
Y_0(s)=e^{-\sqrt{2}s}T_0(p), \quad  p=2\sqrt{2}s, 
\end{equation}
substitution shows that $T(p)$ satisfies Laguerre's equation
\begin{equation} \label{Lag}
p\frac{d^2T_0}{dp^2}+(1-p)\frac{dT_0}{dp}+\frac{1}{2}\left(\frac{\sqrt{2}}{\lambda_0}-1\right)T_0=0.
\end{equation}
Non-singular solutions of the latter, which are consistent with \eqref{IBP conditions} and the required attenuation rate of $Y_0(s)$, exist only when the factor multiplying $T$ in \eqref{Lag} is a non-negative integer. In those cases the solutions are proportional to the Laguerre polynomials. Hence, the leading-order eigenvalues are
\begin{equation} \label{leading}
\lambda_0\ub{n}=\frac{\sqrt{2}}{2n+1}, \quad n=0,1,2,\ldots
\end{equation}
The corresponding transformed solutions are
\begin{equation} \label{Yn}
Y_0\ub{n}(s)=\mathcal{K}\ub{n}e^{-\sqrt{2}s}L_n(2\sqrt{2}s),
\end{equation}
where $\mathcal{K}\ub{n}$ is an arbitrary constant, and $L_n$ is the Legendre polynomial of order $n$.
It is straightforward to invert \eqref{hank 0} to get the eigen-potentials $\psi_0\ub{n}$. For example, noting that $L_0(p)=1$ and $L_1(p)=1-p$, 
the first two modes are
\begin{equation} \label{mode 0}
\psi\ub{0}_0=\frac{\mathcal{K}\ub{0}}{\left[R^2+(\bar{Z}+\sqrt{2})^2\right]^{1/2}}
\end{equation}
and
\begin{equation} \label{mode 1}
\psi\ub{1}_0=\psi\ub{0}_0-\frac{2\sqrt{2}\mathcal{K}\ub{1}(\bar{Z}+\sqrt{2})}{\left[R^2+(\bar{Z}+\sqrt{2})^2\right]^{3/2}}.
\end{equation}

Proceeding to the next logarithmic order, we see that the problem governing $\psi\ub{n}_1$ is similar to the one governing $\psi\ub{n}_0$, only that the homogeneous boundary condition \eqref{bc 0} is replaced by an inhomogeneous one. Dropping the $(n)$ superscript for the moment, the latter condition follows from \eqref{reduced bc} as
\begin{equation}\label{bc 1}
\psi_1 +\lambda_0H(R)\pd{\psi_1}{\bar{Z}}=\frac{\lambda_1}{\lambda_0}\psi_0+\int_0^{\infty}\frac{\psi_0R}{H(R)}\,dR. 
\end{equation}
From the preceding analysis it is clear that, when $\lambda_0$ is given by \eqref{leading}, the problem governing $\psi_1$ possesses a nontrivial homogeneous solution. Hence, the Fredholm alternative suggests a solvability condition on the forcing terms in \eqref{bc 1}. Indeed, by subtracting \eqref{bc 1} multiplied by $R\psi_0/H$ from \eqref{bc 0} multiplied by $R\psi_1/H$, followed by integration with respect to $R$ and use of Green's second identity, we find
\begin{equation} \label{solv}
\lambda_1=-\lambda_0{\left(\int_0^{\infty}\frac{\psi_0R}{H(R)}\,dR\right)^2}\Big{/}{\int_0^{\infty}\frac{\psi_0^2R}{H(R)}\,dR}.
\end{equation}
The integrals in \eqref{solv} can be readily evaluated by use of Parseval's identity and the inversion-symmetry of the Hankel transform, together with the orthonormality of the Laguerre polynomials, see \eqref{lag int 1}. One finds the simple result $\lambda_1=-2\sqrt{2}\left(\lambda_0\right)^2$, 
whereby reinstating the $(n)$ superscript gives the two-term logarithmic expansion
\begin{equation} \label{two terms}
\lambda\ub{n}\sim \frac{\sqrt{2}}{2n+1}\left(1-\frac{4}{2n+1}\frac{1}{\ln(1/h)}+\cdots \right).
\end{equation}

\subsection{Algebraically accurate result}
\label{ssec:algebra}
It is clear that the logarithmic error in \eqref{two terms} limits the usefulness of that result to exceedingly small $h$. To improve on this, we now return to the reduced problem of \S\ref{ssec:reduced} and attempt a complete solution for the $\lambda$ eigenvalues, treating logarithmic terms in par with $O(1)$ terms. This is equivalent to calculating all of the terms in expansion \eqref{log exp}. We begin by writing the solution as [cf.~\eqref{hank 0}, \eqref{hank def} and \eqref{psi 0}]
\begin{equation}\label{Y algebra}
\psi=\mathcal{H}^{-1}[\hat\psi],  \,\,\text{where}\,\, \hat{\psi}=\frac{1}{s}{Y(s)}e^{-s\bar{Z}}.
\end{equation}
Applying the Hankel transform to \eqref{reduced bc}, we find 
\begin{equation} \label{Y eq}
\frac{1}{s}\frac{d}{ds}\left(s\frac{dY}{ds}\right)+2\left(\frac{1}{\lambda s }-1\right)Y=-\frac{4\mathcal{V}}{\lambda}\frac{\delta(s)}{s},
\end{equation}
where $\delta(s)$ is the Dirac delta function. Eq.~\eqref{Y eq} is to be interpreted in the sense of distributions \cite{Crighton:12}, which is easier to do when identifying the Hankel transform as a two-dimensional Fourier transform of a radially symmetric function, $s$ being the radial distance in Fourier-space. Crucially, conditions \eqref{IBP conditions} are now inapplicable; rather, balancing singularities at the origin yields the condition 
\begin{equation} \label{ln condition}
Y(s)\sim-\frac{2\mathcal{V}}{\lambda}\ln s  \quad \text{as} \quad s\to0.
\end{equation}

Following \eqref{Y to T}, we set $Y(s)=e^{-\sqrt{2}s}T(p)$, where $p=2\sqrt{2}s$. This gives [cf.~\eqref{Lag}]
\begin{equation} \label{algebra T}
p\frac{d^2T}{dp^2}+(1-p)\frac{dT}{dp}+\tilde{n}T=0, \quad p>0,
\end{equation}
where $\tilde{n}$ is defined through
\begin{equation}\label{n tilde}
\lambda=\frac{\sqrt{2}}{2\tilde{n}+1}.
\end{equation}
We arrive at an eigenvalue problem consisting of \eqref{algebra T}, the ``boundary'' condition
\begin{equation} \label{ln condition T}
T(p)\sim -\frac{2\mathcal{V}}{\lambda}\ln p \quad \text{as} \quad p\to0,
\end{equation}
and the condition that $T$ does not grow too fast as $p\to\infty$. 
The solutions found in \S\ref{ssec:logarithmic}, which correspond to non-negative integer $\tilde{n}$ and $\mathcal{V}=0$, are  inconsistent with condition \eqref{ln condition T}. Hence, in the present scheme, where logarithmic errors are not tolerated, $\mathcal{V}\ne0$. In the latter case, excluding non-negative integer $\tilde{n}$, solutions of \eqref{algebra T} which are logarithmically singular as $p\to0$ and have acceptable behaviour at large $p$ are proportional to $U(-\tilde{n},1,p)$, where $U$ is the confluent hypergeometric function of the second kind \cite{Abramowitz:book}. 
Noting that
\begin{equation}\label{U asymp}
U(-\tilde{n},1,p)\sim -\frac{1}{\Gamma(-\tilde{n})}\left[\ln p + \Psi(-\tilde{n}) + 2\gamma_e\right]+o(1)
\end{equation}
as $p\to0$, where $\Gamma(x)$ is the Gamma function and $\Psi(x)=\Gamma'(x)/\Gamma(x)$ is the Digamma function, condition \eqref{ln condition T} is satisfied by
\begin{equation} \label{T}
T(p)=\frac{2\mathcal{V}}{\lambda}\Gamma(-\tilde{n})U(-\tilde{n},1,p).
\end{equation}

It remains to satisfy the global condition, that is, to relate $\mathcal{V}$ to the right hand side of \eqref{reduced bc} [cf.~\eqref{V eq}]. It is convenient to first rewrite that condition as
\begin{equation}\label{condition int}
\ln\frac{2}{h}+2\gamma_e=\int_0^{\infty}\left(\lambda\pd{}{\bar{Z}}\left(\psi/\mathcal{V}\right)+\frac{1}{H}\right)R\,dR.
\end{equation}
The integral on the right can be evaluated by using Parseval's identity. Noting that the Hankel transform of $1/H$ is $2K_0(s\sqrt{2})$, $K_0$ being the modified Bessel function of the second kind, that integral is
\begin{equation}\label{the integral}
2\int_0^{\infty}\left[-\frac{\lambda}{\mathcal{V}} Y(s)+2K_0\left(s\sqrt{2}\right)\right]\delta(s)\,ds.
\end{equation}
Given \eqref{U asymp}, and since $K_0(x)\sim -\ln(x/2)-\gamma_e$ as $x\to0$, the apparent logarithmic singularity of the bracketed terms cancels, whereby the sifting property of the Dirac delta function can be effected to show that \eqref{the integral}  $=2\ln 4 +2\Psi(-\tilde{n})+2\gamma_e$. Substitution of this result into \eqref{condition int} furnishes a transcendental equation for $\tilde{n}$:
\begin{equation} \label{trans}
\quad 2\Psi(-\tilde{n})=\ln\frac{1}{8h}.
\end{equation}
Since $\Psi(x)$ diverges at non-positive integer $x$, it is easy to see that \eqref{trans} has multiple non-integer solutions $\tilde{n}$ which (very slowly) approach $n=0,1,2,$ as $h\to0$ \footnote{Eq.~\eqref{trans} produces one additional negative solution. The latter branch, however, approaches $-\infty$ as $h\to0$, whereby \eqref{algebra result} shows that $\lambda$ becomes small and asymptoticness is lost. Consistency requires disregarding this spurious mode, in agreement with general considerations}. 

\begin{figure}[t]
   \centering
   \includegraphics[width=\columnwidth]{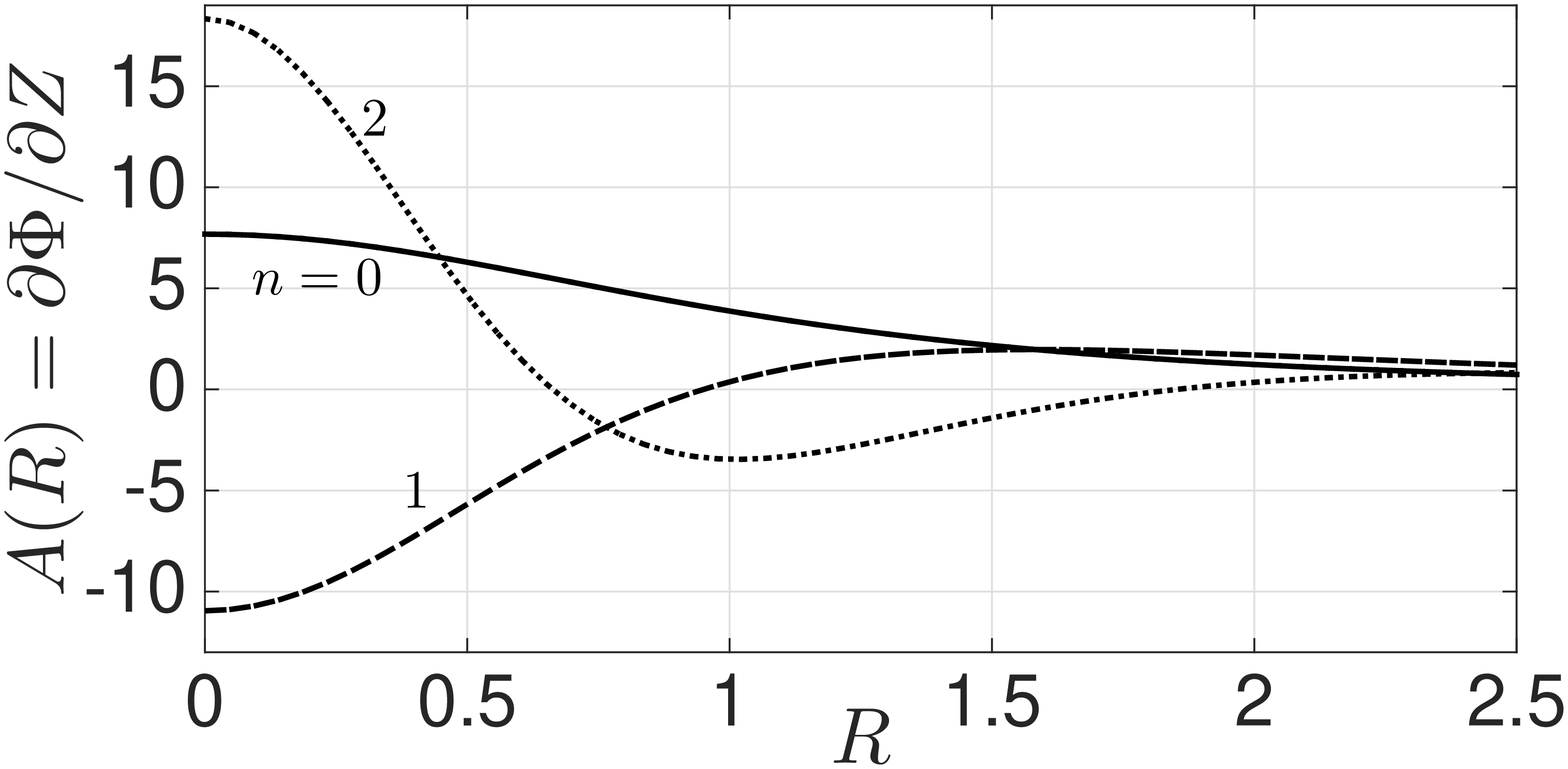} 
   \caption{The radial distribution of the dimensionless transverse field in the gap for the longitudinal modes $n=0,1$ and $2$, for $h=0.001$ and $\mathcal{V}=1$.}
   \label{fig:gapfield}
\end{figure}

Recapitulating, definition \eqref{n tilde} furnishes the eigenvalues as
\begin{equation}\label{algebra result}
\lambda\ub{n}=\frac{\sqrt{2}}{2\tilde{n}(h)+1}, 
\end{equation}
where $\tilde{n}$ is the solution of \eqref{trans} which approaches $n=0,1,2,\ldots$ as $h\to0$. We reiterate that \eqref{algebra result} is algebraically accurate, \textit{viz.}, together with \eqref{eps scaling} it provides the resonance values of the relative dielectric function up to a relative error which scales with some power of $h$. In particular, \eqref{algebra result} is far more accurate then \eqref{two terms}. We note that it is straightforward to recover \eqref{two terms} by expanding $\Psi$ in \eqref{trans} about its poles. Higher order logarithmic terms can be derived in a similar manner. Inverting $\hat\psi$ provides the distribution of the eigenpotentials in the pole to algebraic order [cf.~\eqref{mode 0} and \eqref{mode 1}]. The dominantly transverse field in the gap, which is a function of $R$ alone, is then obtained (up to a minus sign) as $\partial\Phi/\partial Z = A(R)=\psi(R,0)/H(R)$. The radial gap-field distributions of the first three modes are plotted in Fig.~\eqref{fig:gapfield}. 

\section{Induced-dipole moment and field enhancement}
\label{sec:forced}
\subsection{Resonance conditions}
We now return to the excitation problem formulated in \S\ref{sec:formulation}, where the far-field condition \eqref{far} applies, and $\epsilon$ is a complex-valued function of frequency. Whereas a negative real $\epsilon$ is unphysical, the reduced Drude model \eqref{Drude} implies that $-\mathrm{Re}[\epsilon]\gg\mathrm{Im}[\epsilon]$ as long as $\omega\gg\gamma$. Resonance frequencies are safe within this range: even for an extremely narrow gap, say $h=10^{-4}$, Eqs.~\eqref{Drude} and \eqref{eps scaling} suggest $\omega\approx0.1\omega_p$ --- still quite larger than $\gamma$ for silver and gold. Thus, towards studying the damped near-field resonance of the system, it is useful to consider a resonance scenario 
\begin{equation} \label{eps exp}
\epsilon\sim  \epsilon\ub{n}+i\epsilon_i, \quad \epsilon_i\ll h^{-1/2}.
\end{equation}
Here, for a given $n=0,1,2,\ldots$, $\epsilon\ub{n}$ is a resonance value $\sim -\lambda\ub{n}(h)h^{-1/2}+O(1)$, where $\lambda\ub{n}=O(1)$ is given by \eqref{algebra result}, and $i\epsilon_i$ is the imaginary deviation from the resonance values. 

Given  \eqref{eps exp}, it turns out that an asymptotic description of the field distribution --- and in particular, closed-form approximations for the induced-dipole moment $\mu$ and field enhancement in the gap $G$ --- can be obtained based on the LSP modes calculated in \S\ref{sec:resonant}. The crucial observation is that the nth LSP mode is ``excited'' under \eqref{eps exp}, \textit{viz.}, the leading-order potential is of the form calculated in \S\ref{sec:resonant}, but with an asymptotically large multiplicative factor. Note that if the potential was not $\gg1$, \eqref{eps exp} would imply a contradictory leading-order problem which, on the one hand, possesses a non-trivial homogeneous solution, and on the other, is forced at large distances by the incident field. Thus, the integral relation \eqref{work mod} can be written, asymptotically, as
\begin{equation}
4\pi\mu\sim-i\epsilon_i\int_{\text{metal}}\bnabla\varphi \bcdot \bnabla\varphi\,dV. \label{work asym}
\end{equation}
We can now substitute our solutions from \S\ref{sec:resonant} (allowing for a large pre-factor) into \eqref{work asym}, and solve for the amplification factor (say, in terms of $\mu$). We note that, in contrast to \eqref{work asym},  the work-dissipation relation \eqref{work} is ambiguous with respect to phase and hence cannot be employed towards making quantitative assessments without further information. We once more precede a detailed analysis with scaling arguments.

\subsection{Scaling arguments}
\label{ssec:scaling2}
Let $\varphi_g$ denote the order-of-magnitude of the potential in the gap region. The asymptotic structure of the potential then follows from the description provided in \S\ref{sec:resonant}. In particular, the transverse field in the gap is $G=O(\varphi_g/h)$,  the potential in the pole region is $O(\varphi_g)$, and on the particle scale the sphere is approximately equipotential, with $\mathcal{V}= O(\varphi_g/\ln(1/h))$, see \eqref{V eq}. Note that \eqref{muV} implies $\mu=O(\mathcal{V})$. Consider now the integral on the right hand side of \eqref{work asym} [the work-dissipation relation \eqref{work} can also be employed]. The dominant contribution arises from the pole domains, where the potential varies rapidly. Indeed, the volume of the pole domains is $O(h^{3/2})$ and the integrand there is $O(\varphi^2_g /h)$, leading to an $O(\varphi_g^2 h^{1/2})$ estimate. The contribution from the approximately equipotential $O(1)$ volume is $O(\varphi_g^2 h)$ at most. We thus find from \eqref{work asym} that $\varphi_g=O(h^{-1/2}/(\epsilon_i \, \ln(1/h)))$, from which estimates for $\mu$ and $G$ readily follow as
\begin{equation} \label{mu G scalings}
\mu \sim O\left(\frac{1}{\epsilon_i\,h^{1/2}\ln^2\frac{1}{h}}\right), \quad G \sim O\left(\frac{1}{\epsilon_i\,h^{3/2}\ln\frac{1}{h}}\right).
\end{equation}

\subsection{Asymptotic formulas}
To go beyond scaling arguments, we use Green's first identity together with \eqref{constraint} to rewrite \eqref{work asym} as
\begin{equation} \label{work asym psi}
\mu\sim-i\epsilon_i h^{1/2} \int_0^{\infty}\left(\psi\pd{\psi}{\bar{Z}}\right)_{\bar Z=0}R\,dR,
\end{equation}
where $\psi$ denotes the deviation of the potential in the ($z>0$) pole domain from $\mathcal{V}$, see \S\ref{sec:resonant}. 
Applying Parseval's theorem to the right-hand-side of \eqref{work asym}, it becomes [cf.~\eqref{Y algebra}]
\begin{equation} \label{right work}
\sim i \epsilon_i h^{1/2}\int_0^{\infty}Y^2(s)\,ds.
\end{equation}
Upon substitution of $Y$ (see below), and noting that $\mu\sim \pi^2\mathcal{V}/{3}$ [cf.~\eqref{muV}], \eqref{work asym psi} can be solved for $\mathcal{V}$, and hence $\mu$. The field enhancement we can then be obtained from
\begin{equation} \label{G in Phi}
G\sim -\frac{1}{h}\pd{\Phi}{Z},
\end{equation}
where $\Phi$ is the gap potential [cf.~\eqref{Phi exp}], and the derivative is evaluated at $R,Z=0$. Using \eqref{pole conditions}, $G$ can also be written as
\begin{equation} \label{G in psi}
\sim \frac{\lambda}{h}\pd{\psi}{\bar Z},
\end{equation}
where the derivative is evaluated at $R=0,\bar{Z}=0$, or in terms of $Y(s)$,
\begin{equation}\label{G in Y}
\sim -\frac{\lambda}{h}\int_0^{\infty}sY(s)\,ds.
\end{equation}

The calculation can be performed at two levels, depending on whether the logarithmically or algebraically accurate expressions for $Y$ and $\lambda$ are employed in the above formulas. We start with the former case, substituting $Y_0(s)$ for $Y$ (see \eqref{Yn}) and $\lambda_0$ for $\lambda$ (see \eqref{leading}). Noting \eqref{V eq}, the arbitrary constant $\mathcal{K}$ in $Y_0$ is to within a logarithmic relative error $\mathcal{K}\sim \mathcal{V}\ln\frac{1}{h}/\lambda_0$. Together with the integrals \eqref{lag int 1} and \eqref{lag int 2}, we find the explicit closed-form expressions
\begin{equation} \label{mu log}
\mu\sim \frac{4\sqrt{2}\pi^4}{9(2n+1)^2i\epsilon_i\,h^{1/2}\ln^2(1/h)}
\end{equation}
and
\begin{equation} \label{G log}
G\sim \frac{(-1)^{n+1}2\sqrt{2}\pi^2}{3(2n+1)i\epsilon_i\,h^{3/2}\ln(1/h)}.
\end{equation}
Expressions \eqref{mu log} and \eqref{G log} have the advantage of being simple, but since they entail an $O(1/\ln(1/h))$ error, they are hardly accurate, as we shall later demonstrate. We note the agreement with the scaling results \eqref{mu G scalings}, and that the excited resonant mode is imaginary, i.e.~out of phase with the incident field. 

To obtain algebraically accurate formulas, we substitute \eqref{T} for $Y$ and \eqref{algebra result} for $\lambda$. Making use of the integrals \eqref{U int 1} and \eqref{U int 2}, the solution scheme described above yields 
\begin{equation} \label{mu algebra}
\mu\sim \frac{\sqrt{2}\pi^4}{9i\epsilon_i h^{1/2}(2\tilde{n}+1)^2\Psi'(-\tilde{n})}
\end{equation}
and
\begin{equation} \label{G algebra}
G\sim -\frac{\sqrt{2}\pi^2}{12i\epsilon_i h^{3/2}}\frac{\Gamma(-\tilde{n})\,_2F_1(2,2,2-\tilde{n},1/2)}{(2\tilde{n}+1)^2\Psi'(-\tilde{n})\Gamma(2-\tilde{n})},
\end{equation}
where $\Psi'(x)$ is the TriGamma function, the derivative of the DiGamma function, $\,_2F_1$ is the hypergeometric function, 
and $\tilde{n}$ is as before the solution of \eqref{trans} that approaches $n=0,1,2,\ldots$ as $h\to0$. We note that \eqref{mu log} and \eqref{G log} can be derived as leading approximations of \eqref{mu algebra} and \eqref{G algebra}, respectively, by noting that $\tilde{n}-n\sim 2/\ln(1/h)$ as $h\to0$, and that $\Psi'(-\tilde{n}) \sim -(\tilde{n}-n)^{-2}$, $\,_2F_1(2,2,2-\tilde{n},1/2)/\Gamma(2-\tilde{n})\sim4 n!(2n+1)$ and $\Gamma(-\tilde{n})\sim (-1)^{n+1}/[n!(\tilde{n}-n)]$ as $\tilde{n}\to n$. 
\begin{figure}[b]
   \centering
      \includegraphics[width=\columnwidth]{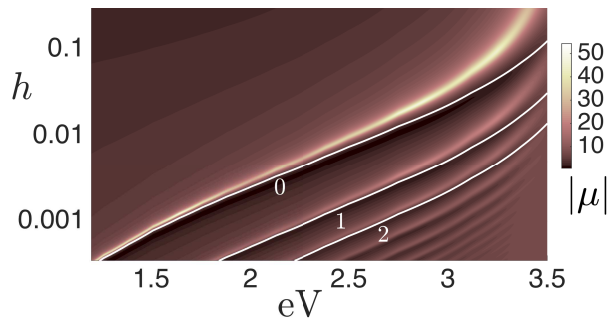} 
   \caption{Absolute magnitude of the dimensionless induced-dipole moment $\mu$ as a function of energy $\hbar\omega$ and $h$, half the dimensionless gap width. The coloured contours represent the semi-numerical solution of the quasi-static problem. The enumerated white lines depict the resonance frequencies of the first three modes as predicted by the asymptotic formula \eqref{algebra result}, together with \eqref{eps scaling}. We employ the empirical silver data of Johnson and Christy \cite{Johnson:72}. }
   \label{fig:res}
\end{figure}

\section{Comparison with semi-numerical solution}
\label{sec:num}
We have validated our asymptotic formulas against a semi-numerical solution of the quasi-static problem. We employ a scheme which is based on the separability of Laplace's equation in bi-spherical coordinates. Briefly, it consists of writing down a formal solution in the form of two infinite series, and determining the coefficients by numerically solving a truncated tri-diagonal system of coupled algebraic equations. The relevant formulas are available elsewhere \cite{Goyette:76,Ruppin:89,Klimov:14}. 

\begin{figure}[t]
   \centering
   \includegraphics[width=\columnwidth]{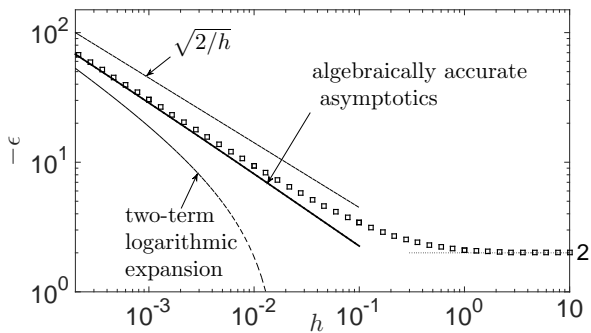} 
   \caption{Zeroth-mode $\epsilon$-eigenvalue as a function of $h$, half the dimensionless gap width. Results of a semi-numerical solution (symbols) are compared with the first term and two terms of the logarithmic expansion \eqref{two terms}, and the algebraically accurate approximation \eqref{algebra result}, cf.~\eqref{eps scaling}. Note also the approach of $\epsilon$ to $-2$ at large $h$. }
   \label{fig:compareres}
\end{figure}

An example calculation is shown in Fig.~\ref{fig:res}, where we have employed the empirical $\epsilon$-data for silver provided by Johnson and Cristy \cite{Johnson:72}. Fig.~\ref{fig:res} shows the absolute magnitude of the induced-dipole moment $\mu$ as a function of the dimensional energy $\hbar\omega$ in $\mathrm{eV}$ units, for a range of $h$. We observe magnitude peaks which, as expected, redshift as $h\to0$. The white lines in Fig.~\ref{fig:res} depict the first three resonance frequencies as predicted from the asymptotic formula \eqref{algebra result} in conjunction with the data of Johnson and Cristy. For small $h$ the latter lines successfully trace the amplitude peaks. Detailed results for the zeroth-mode $\epsilon$-eigenvalues are shown in Fig.~\ref{fig:compareres}. It depicts as a function of $h$ the numerically obtained eigenvalues (symbols) along with the predictions of \eqref{algebra result} and the two-term expansion \eqref{two terms}. The asymptoticness of \eqref{algebra result} as $h\to0$ is evident, with good agreement for $h\lesssim 0.02$ (i.e.~radius-to-gap ratio of $\approx 25$). As expected, the logarithmically accurate approximations are quite poor. Note the approach at large $h$ of $\epsilon$ to $-2$, the Fr\"ohlich value of isolated spheres.

\begin{figure}[t]
   \centering
   \includegraphics[width=\columnwidth]{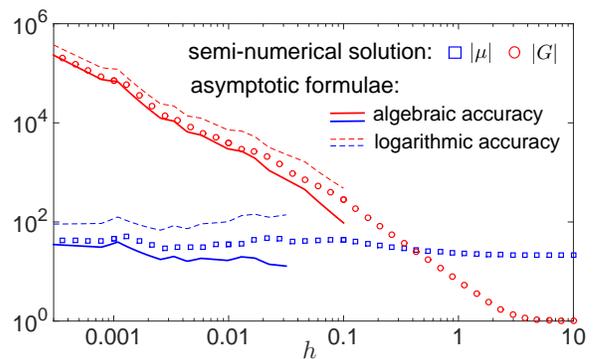} 
   \caption{Numerical peak values of the dimensionless induced-dipole moment $\mu$ and the field enhancement in the gap $G$ at zeroth-mode resonance, as a function of $h$, half the dimensionless gap width. Also shown are the logarithmically accurate approximations \eqref{mu log} and \eqref{G log}, and the algebraically accurate approximations \eqref{mu algebra} and \eqref{G algebra}. We employ the empirical silver data of Johnson and Christy \cite{Johnson:72}.}
   \label{fig:compare}
\end{figure}

Consider next the values of $\mu$ and $G$. Focusing again on the zeroth-mode, we have recorded for a range of $h$ the numerical values attained by $\mu$ and $G$ at their zeroth-mode peaks. The absolute magnitude of these values (which are predominantly imaginary at resonance) are depicted by the symbols in Fig.~\ref{fig:compare}. 
The latter numerical values are compared with the logarithmically accurate asymptotic approximations \eqref{mu log} and \eqref{G log} --- dashed lines, and the algebraically accurate approximations \eqref{mu algebra} and \eqref{G algebra} --- thick lines. Again we see that the latter approximations are rather good, whereas the former are poor for realistically small $h$. 

\section{Discussion}
\label{sec:discussion}
We have employed singular perturbation methods towards analyzing the longitudinal surface modes of a pair of metal spheres in near contact, and the quasi-static excitation of these by an electromagnetic plane wave.  Whereas the techniques of matched asymptotic expansions are routine in many areas of research, it appears that their application in the present context is novel. As demonstrated herein, these allow describing limits of physical extremity in a relatively simple and intuitive manner, notably without the prerequisite of an analytically cumbersome or computationally heavy ``exact'' solution. In particular, the present analysis has furnished accurate asymptotic expressions for the resonance values of the dielectric function (or frequencies) --- see \eqref{algebra result} and \eqref{eps scaling}, and the values at resonance of the induced dipole moment and field-enhancement in the gap --- see \eqref{mu algebra} and \eqref{G algebra}. These expressions have been validated in \S\ref{sec:num} against a semi-numerical solution of the quasi-static problem. 

The above-mentioned expressions are asymptotic in the limit $h\to0$, with algebraic accuracy. That is, they involve a relative error on the order of some power of $h$. These results were preceded with simpler expressions --- see \eqref{two terms}, \eqref{mu log} and \eqref{G log}, which incur however a much larger relative error on the order of $1/\ln(1/h)$. As the comparison in \S\ref{sec:num} unsurprisingly suggests, the latter logarithmic expressions are far less accurate for reasonably small $h$. Nevertheless, the analysis leading to these revealed the essential structure of the longitudinal LSP modes, thereby laying a path towards their algebraic counterparts. We note that similar logarithmic expressions have been derived by approximate manipulations of implicit series solutions in bi-spherical coordinates \cite{Lebedev:13,Klimov:07}. The limited value of these expressions, however, was not emphasised in those works. To be specific, the leading term in \eqref{two terms} agrees with both references, and \eqref{G log} agrees with the expression given in \cite{Lebedev:13}. The logarithmic correction for $\epsilon$ given in Ref.~\cite{Lebedev:13}, however, appears to posses a factor-$2$ error when compared with the second term in \eqref{two terms}. 

In addition to furnishing closed-form approximations, our analysis offers insight into the physical and geometric circumstances enabling a strong frequency redshift and field enhancement. The former has to do largely with the gap morphology, which enables stronger field localization in the dielectric than in the metal --- a prerequisite for large-$|\epsilon|$ eigenvalues. In fact, the scaling of $\epsilon$ with $1/h^{1/2}$ was shown in \S\ref{ssec:scaling} to follow from the particular imbalance in transverse localization pertinent to a locally parabolic boundary. This scaling should therefore apply quite generally to the large-$|\epsilon|$ resonances of essentially arbitrary plasmonic two- and three- dimensional clusters in near contact, or to particles in near contact with a substrate (see Ref.~\cite{Lebedev:10} for an alternative intuitive argument). The origin of field enhancement is more subtle.  The amplitude of an excited mode, represented e.g.~by the configuration-scale induced-dipole moment, is  determined by a balance of work and dissipation, which is in turn affected by the geometrically enabled localization. Disregarding a weak logarithmic dependence upon $h$, the latter enhancement is on the order of $|\epsilon/\epsilon_i|$, which in the near-contact limit is $O(h^{-1/2}/\epsilon_i)$. 
The giant $O(h^{-3/2}/\epsilon_i)$ enhancement in the gap represents a further $O(1/h)$ geometric amplification relative to the induced particle-scale field. We note that in the corresponding planar problem of a nanowire dimer, where scaling arguments similar to those employed herein reveal that the longitudinal modes are determined solely from the gap morphology, the enhancement is smaller, $O(h^{-1}/\epsilon_i)$ \cite{Vorobev:10}. This fundamental difference is apparently overlooked in Ref.~\onlinecite{Klimov:14}, where in the case of a spherical pair the resonant field enhancement in the gap is estimated incorrectly on pg.~278 [cf.~\eqref{mu G scalings}]. 

For the sake of demonstration, we have focused on the axisymmetric longitudinal modes of the spherical dimer configuration, which strongly redshift with vanishing separations and are the ones excited by a field polarized along the line of sphere centers. In many applications the latter modes play the key role,  but this is not always the case, and it is therefore desirable to apply the present approach to LSP modes having different symmetries \cite{Klimov:07}. For example, it has been shown that non-axisymmetric modes of high azimuthal order provide the dominant contribution to van der Waals (vdW) energies at small separations \cite{Klimov:09}. In Ref.~\onlinecite{Klimov:09}, approximate expressions for the vdW energies of sphere dimers and a sphere near a wall were derived by reduction of implicit infinite-series solutions in the double limit of small separations and high azimuthal mode number. While, as already discussed, similar methods gave in Ref.~\onlinecite{Klimov:07} only logarithmic accuracy in the case of longitudinal modes, the dominant modes in the vdW calculation are more highly localized in the gap and accordingly do not involve a logarithmic coupling with the particle-scale field. 

Our methodology can be applied to study various other plasmonic nano-structures that are characterised by multiple length scales. In particular we note the related particle-substrate configuration, which can be addressed along the lines of the present analysis, the substantially simpler two-dimensional problem of a nanowire dimer, and elongated particles. Further desirable generalisations are concerned with physical modelling. Our analysis originates from a quasi-static formulation, and focuses on the near-field response of a small nano-metric particle (still large compared with the gap width). Given our expressions for the induced-dipole moment, estimates for the optical cross section can be obtained via the usual connection formulas \cite{Landau:Book,Bohren:Book,Maier:07}. Ad hoc techniques for improving and extending such estimates to larger nano-structures, wherein the quasi-static induced-dipole is renormalised to account for radiation damping, have proven quite useful \cite{Aubry:10,Le:13}. More rigorous however would be to connect the near and far fields through systematic application of matched asymptotic expansions \cite{Crighton:12,Ando:15}, taking special account of the largeness of $|\epsilon|$, the largeness of the near-field at resonance, and the multiple-scale geometry of the metallic structure; strictly speaking, each of these disparities entails slaving a distinct small perturbation parameter to the Helmholtz parameter $a\omega/c$. 

Probably the most important direction is to incorporate into the present study non-classical physics, which at sub-nanometric separations ultimately suppress the frequency redshift and field enhancement \cite{Fernandez:12,Luo:13,luo:14}. In the framework of the hydrodynamic Drude model \cite{Ciraci:13,Raza:15},  my co-workers and I recently showed \cite{Schnitzer:15arxivB} that the near-contact asymptotics predicted by the latter ``nonlocal'' model constitute a renormalisation of those predicted by the classical ``local'' model considered herein. This finding greatly extends the utility of the present study. 

\begin{acknowledgments}
I wish to thank Vincenzo Giannini, Richard Craster and Stefan Maier for fruitful discussions on this work.
\end{acknowledgments}

\appendix*
\section{Integrals}
\label{app:special}
We employ the integrals
\begin{gather} \label{lag int 1}
\int_0^{\infty}e^{-p}\left[L_n(p)\right]^2\,dp=1, \\
\label{lag int 2}
\int_0^{\infty}pe^{-p/2}L_n(p)\,dp=4(-1)^n(1+2n), \\ 
\label{U int 1}
\int_0^{\infty}e^{-p}\left[U(-\tilde{n},1,p)\right]^2\,dp=\frac{\Psi'(-\tilde{n})}{\left(\Gamma(-\tilde{n})\right)^2}, \\ 
\label{U int 2}
\int_0^{\infty}pe^{-p/2}U(-\tilde{n},1,p)\,dp=\frac{\,_2F_1(2,2,2-\tilde{n},1/2)}{\Gamma(2-\tilde{n})}, 
\end{gather}
for $n=0,1,2,\ldots$ and non-integer $\tilde{n}>0$ \cite{Wolfram,Olver:Book}.

\bibliography{refs.bib}

\begin{thebibliography}{78}%
\makeatletter
\providecommand \@ifxundefined [1]{%
 \@ifx{#1\undefined}
}%
\providecommand \@ifnum [1]{%
 \ifnum #1\expandafter \@firstoftwo
 \else \expandafter \@secondoftwo
 \fi
}%
\providecommand \@ifx [1]{%
 \ifx #1\expandafter \@firstoftwo
 \else \expandafter \@secondoftwo
 \fi
}%
\providecommand \natexlab [1]{#1}%
\providecommand \enquote  [1]{``#1''}%
\providecommand \bibnamefont  [1]{#1}%
\providecommand \bibfnamefont [1]{#1}%
\providecommand \citenamefont [1]{#1}%
\providecommand \href@noop [0]{\@secondoftwo}%
\providecommand \href [0]{\begingroup \@sanitize@url \@href}%
\providecommand \@href[1]{\@@startlink{#1}\@@href}%
\providecommand \@@href[1]{\endgroup#1\@@endlink}%
\providecommand \@sanitize@url [0]{\catcode `\\12\catcode `\$12\catcode
  `\&12\catcode `\#12\catcode `\^12\catcode `\_12\catcode `\%12\relax}%
\providecommand \@@startlink[1]{}%
\providecommand \@@endlink[0]{}%
\providecommand \url  [0]{\begingroup\@sanitize@url \@url }%
\providecommand \@url [1]{\endgroup\@href {#1}{\urlprefix }}%
\providecommand \urlprefix  [0]{URL }%
\providecommand \Eprint [0]{\href }%
\providecommand \doibase [0]{http://dx.doi.org/}%
\providecommand \selectlanguage [0]{\@gobble}%
\providecommand \bibinfo  [0]{\@secondoftwo}%
\providecommand \bibfield  [0]{\@secondoftwo}%
\providecommand \translation [1]{[#1]}%
\providecommand \BibitemOpen [0]{}%
\providecommand \bibitemStop [0]{}%
\providecommand \bibitemNoStop [0]{.\EOS\space}%
\providecommand \EOS [0]{\spacefactor3000\relax}%
\providecommand \BibitemShut  [1]{\csname bibitem#1\endcsname}%
\let\auto@bib@innerbib\@empty
\bibitem [{\citenamefont {Maier}\ and\ \citenamefont
  {Atwater}(2005)}]{Maier:05}%
  \BibitemOpen
  \bibfield  {author} {\bibinfo {author} {\bibfnamefont {S.~A.}\ \bibnamefont
  {Maier}}\ and\ \bibinfo {author} {\bibfnamefont {H.~A.}\ \bibnamefont
  {Atwater}},\ }\href@noop {} {\bibfield  {journal} {\bibinfo  {journal} {J.
  Appl. Phys.}\ }\textbf {\bibinfo {volume} {98}},\ \bibinfo {pages} {011101}
  (\bibinfo {year} {2005})}\BibitemShut {NoStop}%
\bibitem [{\citenamefont {Maier}(2007)}]{Maier:07}%
  \BibitemOpen
  \bibfield  {author} {\bibinfo {author} {\bibfnamefont {S.~A.}\ \bibnamefont
  {Maier}},\ }\href@noop {} {\emph {\bibinfo {title} {Plasmonics: fundamentals
  and applications}}}\ (\bibinfo  {publisher} {Springer Science \& Business
  Media},\ \bibinfo {year} {2007})\BibitemShut {NoStop}%
\bibitem [{\citenamefont {Klimov}(2014)}]{Klimov:14}%
  \BibitemOpen
  \bibfield  {author} {\bibinfo {author} {\bibfnamefont {V.}~\bibnamefont
  {Klimov}},\ }\href@noop {} {\emph {\bibinfo {title} {Nanoplasmonics}}}\
  (\bibinfo  {publisher} {CRC Press},\ \bibinfo {year} {2014})\BibitemShut
  {NoStop}%
\bibitem [{\citenamefont {S{\"o}nnichsen}\ \emph {et~al.}(2005)\citenamefont
  {S{\"o}nnichsen}, \citenamefont {Reinhard}, \citenamefont {Liphardt},\ and\
  \citenamefont {Alivisatos}}]{Sonnichsen:05}%
  \BibitemOpen
  \bibfield  {author} {\bibinfo {author} {\bibfnamefont {C.}~\bibnamefont
  {S{\"o}nnichsen}}, \bibinfo {author} {\bibfnamefont {B.~M.}\ \bibnamefont
  {Reinhard}}, \bibinfo {author} {\bibfnamefont {J.}~\bibnamefont {Liphardt}},
  \ and\ \bibinfo {author} {\bibfnamefont {A.~P.}\ \bibnamefont {Alivisatos}},\
  }\href@noop {} {\bibfield  {journal} {\bibinfo  {journal} {Nat. Biotechnol.}\
  }\textbf {\bibinfo {volume} {23}},\ \bibinfo {pages} {741} (\bibinfo {year}
  {2005})}\BibitemShut {NoStop}%
\bibitem [{\citenamefont {Sukharev}\ and\ \citenamefont
  {Seideman}(2007)}]{Sukharev:07}%
  \BibitemOpen
  \bibfield  {author} {\bibinfo {author} {\bibfnamefont {M.}~\bibnamefont
  {Sukharev}}\ and\ \bibinfo {author} {\bibfnamefont {T.}~\bibnamefont
  {Seideman}},\ }\href@noop {} {\bibfield  {journal} {\bibinfo  {journal} {J.
  Chem. Phys}\ }\textbf {\bibinfo {volume} {126}},\ \bibinfo {pages} {204702}
  (\bibinfo {year} {2007})}\BibitemShut {NoStop}%
\bibitem [{\citenamefont {Xu}\ \emph {et~al.}(2000)\citenamefont {Xu},
  \citenamefont {Aizpurua}, \citenamefont {Kall},\ and\ \citenamefont
  {Apell}}]{Xu:00}%
  \BibitemOpen
  \bibfield  {author} {\bibinfo {author} {\bibfnamefont {H.}~\bibnamefont
  {Xu}}, \bibinfo {author} {\bibfnamefont {J.}~\bibnamefont {Aizpurua}},
  \bibinfo {author} {\bibfnamefont {M.}~\bibnamefont {Kall}}, \ and\ \bibinfo
  {author} {\bibfnamefont {P.}~\bibnamefont {Apell}},\ }\href@noop {}
  {\bibfield  {journal} {\bibinfo  {journal} {Phys. Rev. E}\ }\textbf {\bibinfo
  {volume} {62}},\ \bibinfo {pages} {4318} (\bibinfo {year}
  {2000})}\BibitemShut {NoStop}%
\bibitem [{\citenamefont {Kneipp}\ \emph {et~al.}(2006)\citenamefont {Kneipp},
  \citenamefont {Moskovits},\ and\ \citenamefont {Kneipp}}]{Kneipp:06}%
  \BibitemOpen
  \bibfield  {author} {\bibinfo {author} {\bibfnamefont {K.}~\bibnamefont
  {Kneipp}}, \bibinfo {author} {\bibfnamefont {M.}~\bibnamefont {Moskovits}}, \
  and\ \bibinfo {author} {\bibfnamefont {H.}~\bibnamefont {Kneipp}},\
  }\href@noop {} {\emph {\bibinfo {title} {Surface-enhanced Raman scattering:
  physics and applications}}},\ Vol.\ \bibinfo {volume} {103}\ (\bibinfo
  {publisher} {Springer Science \& Business Media},\ \bibinfo {year}
  {2006})\BibitemShut {NoStop}%
\bibitem [{\citenamefont {Tabor}\ \emph {et~al.}(2008)\citenamefont {Tabor},
  \citenamefont {Murali}, \citenamefont {Mahmoud},\ and\ \citenamefont
  {El-Sayed}}]{Tabor:08}%
  \BibitemOpen
  \bibfield  {author} {\bibinfo {author} {\bibfnamefont {C.}~\bibnamefont
  {Tabor}}, \bibinfo {author} {\bibfnamefont {R.}~\bibnamefont {Murali}},
  \bibinfo {author} {\bibfnamefont {M.}~\bibnamefont {Mahmoud}}, \ and\
  \bibinfo {author} {\bibfnamefont {M.~A.}\ \bibnamefont {El-Sayed}},\
  }\href@noop {} {\bibfield  {journal} {\bibinfo  {journal} {J. Phys. Chem. A}\
  }\textbf {\bibinfo {volume} {113}},\ \bibinfo {pages} {1946} (\bibinfo {year}
  {2008})}\BibitemShut {NoStop}%
\bibitem [{\citenamefont {Oulton}\ \emph {et~al.}(2009)\citenamefont {Oulton},
  \citenamefont {Sorger}, \citenamefont {Zentgraf}, \citenamefont {Ma},
  \citenamefont {Gladden}, \citenamefont {Dai}, \citenamefont {Bartal},\ and\
  \citenamefont {Zhang}}]{Oulton:09}%
  \BibitemOpen
  \bibfield  {author} {\bibinfo {author} {\bibfnamefont {R.~F.}\ \bibnamefont
  {Oulton}}, \bibinfo {author} {\bibfnamefont {V.~J.}\ \bibnamefont {Sorger}},
  \bibinfo {author} {\bibfnamefont {T.}~\bibnamefont {Zentgraf}}, \bibinfo
  {author} {\bibfnamefont {R.~M.}\ \bibnamefont {Ma}}, \bibinfo {author}
  {\bibfnamefont {C.}~\bibnamefont {Gladden}}, \bibinfo {author} {\bibfnamefont
  {L.}~\bibnamefont {Dai}}, \bibinfo {author} {\bibfnamefont {G.}~\bibnamefont
  {Bartal}}, \ and\ \bibinfo {author} {\bibfnamefont {X.}~\bibnamefont
  {Zhang}},\ }\href@noop {} {\bibfield  {journal} {\bibinfo  {journal}
  {Nature}\ }\textbf {\bibinfo {volume} {461}},\ \bibinfo {pages} {629}
  (\bibinfo {year} {2009})}\BibitemShut {NoStop}%
\bibitem [{\citenamefont {Muhlschlegel}\ \emph {et~al.}(2005)\citenamefont
  {Muhlschlegel}, \citenamefont {Eisler}, \citenamefont {Martin}, \citenamefont
  {Hecht},\ and\ \citenamefont {Pohl}}]{Muhlschlegel:05}%
  \BibitemOpen
  \bibfield  {author} {\bibinfo {author} {\bibfnamefont {P.}~\bibnamefont
  {Muhlschlegel}}, \bibinfo {author} {\bibfnamefont {H.~J.}\ \bibnamefont
  {Eisler}}, \bibinfo {author} {\bibfnamefont {O.~J.~F.}\ \bibnamefont
  {Martin}}, \bibinfo {author} {\bibfnamefont {B.}~\bibnamefont {Hecht}}, \
  and\ \bibinfo {author} {\bibfnamefont {D.~W.}\ \bibnamefont {Pohl}},\
  }\href@noop {} {\bibfield  {journal} {\bibinfo  {journal} {Science}\ }\textbf
  {\bibinfo {volume} {308}},\ \bibinfo {pages} {1607} (\bibinfo {year}
  {2005})}\BibitemShut {NoStop}%
\bibitem [{\citenamefont {Muskens}\ \emph {et~al.}(2007)\citenamefont
  {Muskens}, \citenamefont {Giannini}, \citenamefont {Sanchez-Gil},\ and\
  \citenamefont {Gomez~R.}}]{Muskens:07b}%
  \BibitemOpen
  \bibfield  {author} {\bibinfo {author} {\bibfnamefont {O.~L.}\ \bibnamefont
  {Muskens}}, \bibinfo {author} {\bibfnamefont {V.}~\bibnamefont {Giannini}},
  \bibinfo {author} {\bibfnamefont {J.~A.}\ \bibnamefont {Sanchez-Gil}}, \ and\
  \bibinfo {author} {\bibfnamefont {J.}~\bibnamefont {Gomez~R.}},\ }\href@noop
  {} {\bibfield  {journal} {\bibinfo  {journal} {Nano Lett.}\ }\textbf
  {\bibinfo {volume} {7}},\ \bibinfo {pages} {2871} (\bibinfo {year}
  {2007})}\BibitemShut {NoStop}%
\bibitem [{\citenamefont {Novotny}\ and\ \citenamefont
  {Van~Hulst}(2011)}]{Novotny:11}%
  \BibitemOpen
  \bibfield  {author} {\bibinfo {author} {\bibfnamefont {L.}~\bibnamefont
  {Novotny}}\ and\ \bibinfo {author} {\bibfnamefont {N.}~\bibnamefont
  {Van~Hulst}},\ }\href@noop {} {\bibfield  {journal} {\bibinfo  {journal}
  {Nature Photon.}\ }\textbf {\bibinfo {volume} {5}},\ \bibinfo {pages} {83}
  (\bibinfo {year} {2011})}\BibitemShut {NoStop}%
\bibitem [{\citenamefont {Hubert}\ \emph {et~al.}(2007)\citenamefont {Hubert},
  \citenamefont {Billot}, \citenamefont {Adam}, \citenamefont {Bachelot},
  \citenamefont {Royer}, \citenamefont {Grand}, \citenamefont {Gindre},
  \citenamefont {Dorkenoo},\ and\ \citenamefont {Fort}}]{Hubert:07}%
  \BibitemOpen
  \bibfield  {author} {\bibinfo {author} {\bibfnamefont {C.}~\bibnamefont
  {Hubert}}, \bibinfo {author} {\bibfnamefont {L.}~\bibnamefont {Billot}},
  \bibinfo {author} {\bibfnamefont {P.~M.}\ \bibnamefont {Adam}}, \bibinfo
  {author} {\bibfnamefont {R.}~\bibnamefont {Bachelot}}, \bibinfo {author}
  {\bibfnamefont {P.}~\bibnamefont {Royer}}, \bibinfo {author} {\bibfnamefont
  {J.}~\bibnamefont {Grand}}, \bibinfo {author} {\bibfnamefont
  {D.}~\bibnamefont {Gindre}}, \bibinfo {author} {\bibfnamefont {K.~D.}\
  \bibnamefont {Dorkenoo}}, \ and\ \bibinfo {author} {\bibfnamefont
  {A.}~\bibnamefont {Fort}},\ }\href@noop {} {\bibfield  {journal} {\bibinfo
  {journal} {Appl. Phys. Lett.}\ }\textbf {\bibinfo {volume} {90}},\ \bibinfo
  {pages} {181105} (\bibinfo {year} {2007})}\BibitemShut {NoStop}%
\bibitem [{\citenamefont {Kim}\ \emph {et~al.}(2008)\citenamefont {Kim},
  \citenamefont {Jin}, \citenamefont {Kim}, \citenamefont {Park}, \citenamefont
  {Kim},\ and\ \citenamefont {Kim}}]{Kim:08}%
  \BibitemOpen
  \bibfield  {author} {\bibinfo {author} {\bibfnamefont {A.}~\bibnamefont
  {Kim}}, \bibinfo {author} {\bibfnamefont {J.}~\bibnamefont {Jin}}, \bibinfo
  {author} {\bibfnamefont {Y.~J.}\ \bibnamefont {Kim}}, \bibinfo {author}
  {\bibfnamefont {I.~Y.}\ \bibnamefont {Park}}, \bibinfo {author}
  {\bibfnamefont {Y.}~\bibnamefont {Kim}}, \ and\ \bibinfo {author}
  {\bibfnamefont {S.~W.}\ \bibnamefont {Kim}},\ }\href@noop {} {\bibfield
  {journal} {\bibinfo  {journal} {Nature}\ }\textbf {\bibinfo {volume} {453}},\
  \bibinfo {pages} {757} (\bibinfo {year} {2008})}\BibitemShut {NoStop}%
\bibitem [{\citenamefont {Atwater}\ and\ \citenamefont
  {Polman}(2010)}]{Atwater:10}%
  \BibitemOpen
  \bibfield  {author} {\bibinfo {author} {\bibfnamefont {H.~A.}\ \bibnamefont
  {Atwater}}\ and\ \bibinfo {author} {\bibfnamefont {A.}~\bibnamefont
  {Polman}},\ }\href@noop {} {\bibfield  {journal} {\bibinfo  {journal} {Nat.
  Mater.}\ }\textbf {\bibinfo {volume} {9}},\ \bibinfo {pages} {205} (\bibinfo
  {year} {2010})}\BibitemShut {NoStop}%
\bibitem [{\citenamefont {Bohren}\ and\ \citenamefont
  {Huffman}(2008)}]{Bohren:Book}%
  \BibitemOpen
  \bibfield  {author} {\bibinfo {author} {\bibfnamefont {C.~F.}\ \bibnamefont
  {Bohren}}\ and\ \bibinfo {author} {\bibfnamefont {D.~R.}\ \bibnamefont
  {Huffman}},\ }\href@noop {} {\emph {\bibinfo {title} {Absorption and
  scattering of light by small particles}}}\ (\bibinfo  {publisher} {John Wiley
  \& Sons},\ \bibinfo {year} {2008})\BibitemShut {NoStop}%
\bibitem [{\citenamefont {Sosa}\ \emph {et~al.}(2003)\citenamefont {Sosa},
  \citenamefont {Noguez},\ and\ \citenamefont {Barrera}}]{Sosa:03}%
  \BibitemOpen
  \bibfield  {author} {\bibinfo {author} {\bibfnamefont {I.~O.}\ \bibnamefont
  {Sosa}}, \bibinfo {author} {\bibfnamefont {C.}~\bibnamefont {Noguez}}, \ and\
  \bibinfo {author} {\bibfnamefont {R.~G.}\ \bibnamefont {Barrera}},\
  }\href@noop {} {\bibfield  {journal} {\bibinfo  {journal} {J. Phys. Chem. B}\
  }\textbf {\bibinfo {volume} {107}},\ \bibinfo {pages} {6269} (\bibinfo {year}
  {2003})}\BibitemShut {NoStop}%
\bibitem [{\citenamefont {Myroshnychenko}\ \emph {et~al.}(2008)\citenamefont
  {Myroshnychenko}, \citenamefont {Rodr{\'\i}guez-Fern{\'a}ndez}, \citenamefont
  {Pastoriza-Santos}, \citenamefont {Funston}, \citenamefont {Novo},
  \citenamefont {Mulvaney}, \citenamefont {Liz-Marz{\'a}n},\ and\ \citenamefont
  {de~Abajo}}]{Myroshnychenko:08}%
  \BibitemOpen
  \bibfield  {author} {\bibinfo {author} {\bibfnamefont {V.}~\bibnamefont
  {Myroshnychenko}}, \bibinfo {author} {\bibfnamefont {J.}~\bibnamefont
  {Rodr{\'\i}guez-Fern{\'a}ndez}}, \bibinfo {author} {\bibfnamefont
  {I.}~\bibnamefont {Pastoriza-Santos}}, \bibinfo {author} {\bibfnamefont
  {A.~M.}\ \bibnamefont {Funston}}, \bibinfo {author} {\bibfnamefont
  {C.}~\bibnamefont {Novo}}, \bibinfo {author} {\bibfnamefont {P.}~\bibnamefont
  {Mulvaney}}, \bibinfo {author} {\bibfnamefont {L.~M.}\ \bibnamefont
  {Liz-Marz{\'a}n}}, \ and\ \bibinfo {author} {\bibfnamefont {F.~J.~G.}\
  \bibnamefont {de~Abajo}},\ }\href@noop {} {\bibfield  {journal} {\bibinfo
  {journal} {Chem. Soc. Rev.}\ }\textbf {\bibinfo {volume} {37}},\ \bibinfo
  {pages} {1792} (\bibinfo {year} {2008})}\BibitemShut {NoStop}%
\bibitem [{\citenamefont {Rueting}\ and\ \citenamefont
  {Uecker}(2010)}]{Rueting:10}%
  \BibitemOpen
  \bibfield  {author} {\bibinfo {author} {\bibfnamefont {F.}~\bibnamefont
  {Rueting}}\ and\ \bibinfo {author} {\bibfnamefont {H.}~\bibnamefont
  {Uecker}},\ }\href@noop {} {\bibfield  {journal} {\bibinfo  {journal} {arXiv
  preprint arXiv:1002.4337}\ } (\bibinfo {year} {2010})}\BibitemShut {NoStop}%
\bibitem [{\citenamefont {Chuntonov}\ and\ \citenamefont
  {Haran}(2011)}]{Chuntonov:11}%
  \BibitemOpen
  \bibfield  {author} {\bibinfo {author} {\bibfnamefont {L.}~\bibnamefont
  {Chuntonov}}\ and\ \bibinfo {author} {\bibfnamefont {G.}~\bibnamefont
  {Haran}},\ }\href@noop {} {\bibfield  {journal} {\bibinfo  {journal} {J.
  Phys. Chem. C}\ }\textbf {\bibinfo {volume} {115}},\ \bibinfo {pages} {19488}
  (\bibinfo {year} {2011})}\BibitemShut {NoStop}%
\bibitem [{\citenamefont {Dobrzynski}\ and\ \citenamefont
  {Maradudin}(1972)}]{Dobrzynski:72}%
  \BibitemOpen
  \bibfield  {author} {\bibinfo {author} {\bibfnamefont {L.}~\bibnamefont
  {Dobrzynski}}\ and\ \bibinfo {author} {\bibfnamefont {A.~A.}\ \bibnamefont
  {Maradudin}},\ }\href@noop {} {\bibfield  {journal} {\bibinfo  {journal}
  {Phys. Rev. B}\ }\textbf {\bibinfo {volume} {6}},\ \bibinfo {pages} {3810}
  (\bibinfo {year} {1972})}\BibitemShut {NoStop}%
\bibitem [{\citenamefont {Luo}\ \emph {et~al.}(2012)\citenamefont {Luo},
  \citenamefont {Lei}, \citenamefont {Maier},\ and\ \citenamefont
  {Pendry}}]{luo:12}%
  \BibitemOpen
  \bibfield  {author} {\bibinfo {author} {\bibfnamefont {Y.}~\bibnamefont
  {Luo}}, \bibinfo {author} {\bibfnamefont {D.~Y.}\ \bibnamefont {Lei}},
  \bibinfo {author} {\bibfnamefont {S.~A.}\ \bibnamefont {Maier}}, \ and\
  \bibinfo {author} {\bibfnamefont {J.~B.}\ \bibnamefont {Pendry}},\
  }\href@noop {} {\bibfield  {journal} {\bibinfo  {journal} {Phys. Rev. Lett.}\
  }\textbf {\bibinfo {volume} {108}},\ \bibinfo {pages} {023901} (\bibinfo
  {year} {2012})}\BibitemShut {NoStop}%
\bibitem [{\citenamefont {Aizpurua}\ \emph {et~al.}(2005)\citenamefont
  {Aizpurua}, \citenamefont {Bryant}, \citenamefont {Richter}, \citenamefont
  {De~Abajo}, \citenamefont {Kelley},\ and\ \citenamefont
  {Mallouk}}]{Aizpurua:05}%
  \BibitemOpen
  \bibfield  {author} {\bibinfo {author} {\bibfnamefont {J.}~\bibnamefont
  {Aizpurua}}, \bibinfo {author} {\bibfnamefont {G.~W.}\ \bibnamefont
  {Bryant}}, \bibinfo {author} {\bibfnamefont {L.~J.}\ \bibnamefont {Richter}},
  \bibinfo {author} {\bibfnamefont {F.~J.~G.}\ \bibnamefont {De~Abajo}},
  \bibinfo {author} {\bibfnamefont {B.~K.}\ \bibnamefont {Kelley}}, \ and\
  \bibinfo {author} {\bibfnamefont {T.}~\bibnamefont {Mallouk}},\ }\href@noop
  {} {\bibfield  {journal} {\bibinfo  {journal} {Phys. Rev. B}\ }\textbf
  {\bibinfo {volume} {71}},\ \bibinfo {pages} {235420} (\bibinfo {year}
  {2005})}\BibitemShut {NoStop}%
\bibitem [{\citenamefont {Andersen}\ and\ \citenamefont
  {Rowlen}(2002)}]{Andersen:02}%
  \BibitemOpen
  \bibfield  {author} {\bibinfo {author} {\bibfnamefont {P.~C.}\ \bibnamefont
  {Andersen}}\ and\ \bibinfo {author} {\bibfnamefont {K.~L.}\ \bibnamefont
  {Rowlen}},\ }\href@noop {} {\bibfield  {journal} {\bibinfo  {journal} {Appl.
  Spectrosc.}\ }\textbf {\bibinfo {volume} {56}},\ \bibinfo {pages} {124A}
  (\bibinfo {year} {2002})}\BibitemShut {NoStop}%
\bibitem [{\citenamefont {Guzatov}\ and\ \citenamefont
  {Klimov}(2011)}]{Guzatov:11}%
  \BibitemOpen
  \bibfield  {author} {\bibinfo {author} {\bibfnamefont {D.~V.}\ \bibnamefont
  {Guzatov}}\ and\ \bibinfo {author} {\bibfnamefont {V.~V.}\ \bibnamefont
  {Klimov}},\ }\href@noop {} {\bibfield  {journal} {\bibinfo  {journal} {New J.
  Phys.}\ }\textbf {\bibinfo {volume} {13}},\ \bibinfo {pages} {053034}
  (\bibinfo {year} {2011})}\BibitemShut {NoStop}%
\bibitem [{\citenamefont {Wang}\ \emph {et~al.}(2013)\citenamefont {Wang},
  \citenamefont {Plouraboue},\ and\ \citenamefont {Chang}}]{Wang:13}%
  \BibitemOpen
  \bibfield  {author} {\bibinfo {author} {\bibfnamefont {Y.}~\bibnamefont
  {Wang}}, \bibinfo {author} {\bibfnamefont {F.}~\bibnamefont {Plouraboue}}, \
  and\ \bibinfo {author} {\bibfnamefont {H.-C.}\ \bibnamefont {Chang}},\
  }\href@noop {} {\bibfield  {journal} {\bibinfo  {journal} {Opt. Express}\
  }\textbf {\bibinfo {volume} {21}},\ \bibinfo {pages} {6609} (\bibinfo {year}
  {2013})}\BibitemShut {NoStop}%
\bibitem [{\citenamefont {Aubry}\ \emph
  {et~al.}(2010{\natexlab{a}})\citenamefont {Aubry}, \citenamefont {Lei},
  \citenamefont {Maier},\ and\ \citenamefont {Pendry}}]{Aubry:10A}%
  \BibitemOpen
  \bibfield  {author} {\bibinfo {author} {\bibfnamefont {A.}~\bibnamefont
  {Aubry}}, \bibinfo {author} {\bibfnamefont {D.~Y.}\ \bibnamefont {Lei}},
  \bibinfo {author} {\bibfnamefont {S.~A.}\ \bibnamefont {Maier}}, \ and\
  \bibinfo {author} {\bibfnamefont {J.~B.}\ \bibnamefont {Pendry}},\
  }\href@noop {} {\bibfield  {journal} {\bibinfo  {journal} {Phys. Rev. Lett.}\
  }\textbf {\bibinfo {volume} {105}},\ \bibinfo {pages} {233901} (\bibinfo
  {year} {2010}{\natexlab{a}})}\BibitemShut {NoStop}%
\bibitem [{\citenamefont {Aubry}\ \emph {et~al.}(2011)\citenamefont {Aubry},
  \citenamefont {Lei}, \citenamefont {Maier},\ and\ \citenamefont
  {Pendry}}]{Aubry:11}%
  \BibitemOpen
  \bibfield  {author} {\bibinfo {author} {\bibfnamefont {A.}~\bibnamefont
  {Aubry}}, \bibinfo {author} {\bibfnamefont {D.~Y.}\ \bibnamefont {Lei}},
  \bibinfo {author} {\bibfnamefont {S.~A.}\ \bibnamefont {Maier}}, \ and\
  \bibinfo {author} {\bibfnamefont {J.~B.}\ \bibnamefont {Pendry}},\
  }\href@noop {} {\bibfield  {journal} {\bibinfo  {journal} {ACS nano}\
  }\textbf {\bibinfo {volume} {5}},\ \bibinfo {pages} {3293} (\bibinfo {year}
  {2011})}\BibitemShut {NoStop}%
\bibitem [{\citenamefont {Luo}\ \emph {et~al.}(2014)\citenamefont {Luo},
  \citenamefont {Zhao},\ and\ \citenamefont {Pendry}}]{luo:14}%
  \BibitemOpen
  \bibfield  {author} {\bibinfo {author} {\bibfnamefont {Y.}~\bibnamefont
  {Luo}}, \bibinfo {author} {\bibfnamefont {R.}~\bibnamefont {Zhao}}, \ and\
  \bibinfo {author} {\bibfnamefont {J.~B.}\ \bibnamefont {Pendry}},\
  }\href@noop {} {\bibfield  {journal} {\bibinfo  {journal} {Proc. Natl. Acad.
  Sci. U.S.A.}\ }\textbf {\bibinfo {volume} {111}},\ \bibinfo {pages} {18422}
  (\bibinfo {year} {2014})}\BibitemShut {NoStop}%
\bibitem [{\citenamefont {Paley}\ \emph {et~al.}(1993)\citenamefont {Paley},
  \citenamefont {Radchik},\ and\ \citenamefont {Smith}}]{Paley:93}%
  \BibitemOpen
  \bibfield  {author} {\bibinfo {author} {\bibfnamefont {A.~V.}\ \bibnamefont
  {Paley}}, \bibinfo {author} {\bibfnamefont {A.~V.}\ \bibnamefont {Radchik}},
  \ and\ \bibinfo {author} {\bibfnamefont {G.~B.}\ \bibnamefont {Smith}},\
  }\href@noop {} {\bibfield  {journal} {\bibinfo  {journal} {J. Appl. Phys.}\
  }\textbf {\bibinfo {volume} {73}},\ \bibinfo {pages} {3446} (\bibinfo {year}
  {1993})}\BibitemShut {NoStop}%
\bibitem [{\citenamefont {Fern{\'a}ndez-Dom{\'\i}nguez}\ \emph
  {et~al.}(2010)\citenamefont {Fern{\'a}ndez-Dom{\'\i}nguez}, \citenamefont
  {Maier},\ and\ \citenamefont {Pendry}}]{Fernandez:10}%
  \BibitemOpen
  \bibfield  {author} {\bibinfo {author} {\bibfnamefont {A.~I.}\ \bibnamefont
  {Fern{\'a}ndez-Dom{\'\i}nguez}}, \bibinfo {author} {\bibfnamefont {S.~A.}\
  \bibnamefont {Maier}}, \ and\ \bibinfo {author} {\bibfnamefont {J.~B.}\
  \bibnamefont {Pendry}},\ }\href@noop {} {\bibfield  {journal} {\bibinfo
  {journal} {Phys. Rev. Lett.}\ }\textbf {\bibinfo {volume} {105}},\ \bibinfo
  {pages} {266807} (\bibinfo {year} {2010})}\BibitemShut {NoStop}%
\bibitem [{\citenamefont {Jung}\ and\ \citenamefont
  {Pedersen}(2012)}]{Jung:12}%
  \BibitemOpen
  \bibfield  {author} {\bibinfo {author} {\bibfnamefont {J.}~\bibnamefont
  {Jung}}\ and\ \bibinfo {author} {\bibfnamefont {T.~G.}\ \bibnamefont
  {Pedersen}},\ }\href@noop {} {\bibfield  {journal} {\bibinfo  {journal} {J.
  Appl. Phys.}\ }\textbf {\bibinfo {volume} {112}},\ \bibinfo {pages} {064312}
  (\bibinfo {year} {2012})}\BibitemShut {NoStop}%
\bibitem [{\citenamefont {Lei}\ \emph {et~al.}(2012)\citenamefont {Lei},
  \citenamefont {Fernández-Domínguez}, \citenamefont {Sonnefraud},
  \citenamefont {Appavoo}, \citenamefont {Haglund~J.}, \citenamefont {Pendry},\
  and\ \citenamefont {Maier}}]{Lei:12}%
  \BibitemOpen
  \bibfield  {author} {\bibinfo {author} {\bibfnamefont {D.~Y.}\ \bibnamefont
  {Lei}}, \bibinfo {author} {\bibfnamefont {A.~I.}\ \bibnamefont
  {Fernández-Domínguez}}, \bibinfo {author} {\bibfnamefont {Y.}~\bibnamefont
  {Sonnefraud}}, \bibinfo {author} {\bibfnamefont {K.}~\bibnamefont {Appavoo}},
  \bibinfo {author} {\bibfnamefont {R.~F.}\ \bibnamefont {Haglund~J.}},
  \bibinfo {author} {\bibfnamefont {J.~B.}\ \bibnamefont {Pendry}}, \ and\
  \bibinfo {author} {\bibfnamefont {S.~A.}\ \bibnamefont {Maier}},\ }\href@noop
  {} {\bibfield  {journal} {\bibinfo  {journal} {ACS Nano}\ }\textbf {\bibinfo
  {volume} {6}},\ \bibinfo {pages} {1380} (\bibinfo {year} {2012})}\BibitemShut
  {NoStop}%
\bibitem [{\citenamefont {Cirac{\`\i}}\ \emph {et~al.}(2012)\citenamefont
  {Cirac{\`\i}}, \citenamefont {Hill}, \citenamefont {Mock}, \citenamefont
  {Urzhumov}, \citenamefont {Fern{\'a}ndez-Dom{\'\i}nguez}, \citenamefont
  {Maier}, \citenamefont {Pendry}, \citenamefont {Chilkoti},\ and\
  \citenamefont {Smith}}]{Ciraci:12}%
  \BibitemOpen
  \bibfield  {author} {\bibinfo {author} {\bibfnamefont {C.}~\bibnamefont
  {Cirac{\`\i}}}, \bibinfo {author} {\bibfnamefont {R.~T.}\ \bibnamefont
  {Hill}}, \bibinfo {author} {\bibfnamefont {J.~J.}\ \bibnamefont {Mock}},
  \bibinfo {author} {\bibfnamefont {Y.}~\bibnamefont {Urzhumov}}, \bibinfo
  {author} {\bibfnamefont {A.~I.}\ \bibnamefont
  {Fern{\'a}ndez-Dom{\'\i}nguez}}, \bibinfo {author} {\bibfnamefont {S.~A.}\
  \bibnamefont {Maier}}, \bibinfo {author} {\bibfnamefont {J.~B.}\ \bibnamefont
  {Pendry}}, \bibinfo {author} {\bibfnamefont {A.}~\bibnamefont {Chilkoti}}, \
  and\ \bibinfo {author} {\bibfnamefont {D.~R.}\ \bibnamefont {Smith}},\
  }\href@noop {} {\bibfield  {journal} {\bibinfo  {journal} {Science}\ }\textbf
  {\bibinfo {volume} {337}},\ \bibinfo {pages} {1072} (\bibinfo {year}
  {2012})}\BibitemShut {NoStop}%
\bibitem [{\citenamefont {Ruppin}(1982)}]{Ruppin:82}%
  \BibitemOpen
  \bibfield  {author} {\bibinfo {author} {\bibfnamefont {R.}~\bibnamefont
  {Ruppin}},\ }\href@noop {} {\bibfield  {journal} {\bibinfo  {journal} {Phys.
  Rev. B}\ }\textbf {\bibinfo {volume} {26}},\ \bibinfo {pages} {3440}
  (\bibinfo {year} {1982})}\BibitemShut {NoStop}%
\bibitem [{\citenamefont {Mayergoyz}\ \emph {et~al.}(2005)\citenamefont
  {Mayergoyz}, \citenamefont {Fredkin},\ and\ \citenamefont
  {Zhang}}]{Mayergoyz:05}%
  \BibitemOpen
  \bibfield  {author} {\bibinfo {author} {\bibfnamefont {I.~D.}\ \bibnamefont
  {Mayergoyz}}, \bibinfo {author} {\bibfnamefont {D.~R.}\ \bibnamefont
  {Fredkin}}, \ and\ \bibinfo {author} {\bibfnamefont {Z.}~\bibnamefont
  {Zhang}},\ }\href@noop {} {\bibfield  {journal} {\bibinfo  {journal} {Phys.
  Rev. B}\ }\textbf {\bibinfo {volume} {72}},\ \bibinfo {pages} {155412}
  (\bibinfo {year} {2005})}\BibitemShut {NoStop}%
\bibitem [{\citenamefont {Romero}\ \emph {et~al.}(2006)\citenamefont {Romero},
  \citenamefont {Aizpurua}, \citenamefont {Bryant},\ and\ \citenamefont
  {Garc{\'\i}a De~Abajo}}]{Romero:06}%
  \BibitemOpen
  \bibfield  {author} {\bibinfo {author} {\bibfnamefont {I.}~\bibnamefont
  {Romero}}, \bibinfo {author} {\bibfnamefont {J.}~\bibnamefont {Aizpurua}},
  \bibinfo {author} {\bibfnamefont {G.~W.}\ \bibnamefont {Bryant}}, \ and\
  \bibinfo {author} {\bibfnamefont {F.~J.}\ \bibnamefont {Garc{\'\i}a
  De~Abajo}},\ }\href@noop {} {\bibfield  {journal} {\bibinfo  {journal} {Opt.
  Express}\ }\textbf {\bibinfo {volume} {14}},\ \bibinfo {pages} {9988}
  (\bibinfo {year} {2006})}\BibitemShut {NoStop}%
\bibitem [{\citenamefont {Giannini}\ and\ \citenamefont
  {Sanchez-Gil}(2007)}]{Giannini:07}%
  \BibitemOpen
  \bibfield  {author} {\bibinfo {author} {\bibfnamefont {V.}~\bibnamefont
  {Giannini}}\ and\ \bibinfo {author} {\bibfnamefont {J.~A.}\ \bibnamefont
  {Sanchez-Gil}},\ }\href@noop {} {\bibfield  {journal} {\bibinfo  {journal}
  {JOSA A}\ }\textbf {\bibinfo {volume} {24}},\ \bibinfo {pages} {2822}
  (\bibinfo {year} {2007})}\BibitemShut {NoStop}%
\bibitem [{\citenamefont {Encina}\ and\ \citenamefont
  {Coronado}(2010)}]{Encina:10}%
  \BibitemOpen
  \bibfield  {author} {\bibinfo {author} {\bibfnamefont {E.~R.}\ \bibnamefont
  {Encina}}\ and\ \bibinfo {author} {\bibfnamefont {E.~A.}\ \bibnamefont
  {Coronado}},\ }\href@noop {} {\bibfield  {journal} {\bibinfo  {journal} {J.
  Phys. Chem. C}\ }\textbf {\bibinfo {volume} {114}},\ \bibinfo {pages} {3918}
  (\bibinfo {year} {2010})}\BibitemShut {NoStop}%
\bibitem [{\citenamefont {Hohenester}\ and\ \citenamefont
  {Tr{\"u}gler}(2012)}]{Hohenester:12}%
  \BibitemOpen
  \bibfield  {author} {\bibinfo {author} {\bibfnamefont {U.}~\bibnamefont
  {Hohenester}}\ and\ \bibinfo {author} {\bibfnamefont {A.}~\bibnamefont
  {Tr{\"u}gler}},\ }\href@noop {} {\bibfield  {journal} {\bibinfo  {journal}
  {Comput. Phys. Commun.}\ }\textbf {\bibinfo {volume} {183}},\ \bibinfo
  {pages} {370} (\bibinfo {year} {2012})}\BibitemShut {NoStop}%
\bibitem [{\citenamefont {Dutta}\ \emph {et~al.}(2008)\citenamefont {Dutta},
  \citenamefont {Ali}, \citenamefont {Brandl}, \citenamefont {Park},\ and\
  \citenamefont {Nordlander}}]{Dutta:08}%
  \BibitemOpen
  \bibfield  {author} {\bibinfo {author} {\bibfnamefont {C.~M.}\ \bibnamefont
  {Dutta}}, \bibinfo {author} {\bibfnamefont {T.~A.}\ \bibnamefont {Ali}},
  \bibinfo {author} {\bibfnamefont {D.~W.}\ \bibnamefont {Brandl}}, \bibinfo
  {author} {\bibfnamefont {T.~H.}\ \bibnamefont {Park}}, \ and\ \bibinfo
  {author} {\bibfnamefont {P.}~\bibnamefont {Nordlander}},\ }\href@noop {}
  {\bibfield  {journal} {\bibinfo  {journal} {J. Chem. Phys}\ }\textbf
  {\bibinfo {volume} {129}},\ \bibinfo {pages} {084706} (\bibinfo {year}
  {2008})}\BibitemShut {NoStop}%
\bibitem [{\citenamefont {Kraft}\ \emph {et~al.}(2014)\citenamefont {Kraft},
  \citenamefont {Pendry}, \citenamefont {Maier},\ and\ \citenamefont
  {Luo}}]{Kraft:14}%
  \BibitemOpen
  \bibfield  {author} {\bibinfo {author} {\bibfnamefont {M.}~\bibnamefont
  {Kraft}}, \bibinfo {author} {\bibfnamefont {J.~B.}\ \bibnamefont {Pendry}},
  \bibinfo {author} {\bibfnamefont {S.~A.}\ \bibnamefont {Maier}}, \ and\
  \bibinfo {author} {\bibfnamefont {Y.}~\bibnamefont {Luo}},\ }\href@noop {}
  {\bibfield  {journal} {\bibinfo  {journal} {Phys. Rev. B}\ }\textbf {\bibinfo
  {volume} {89}},\ \bibinfo {pages} {245125} (\bibinfo {year}
  {2014})}\BibitemShut {NoStop}%
\bibitem [{\citenamefont {Pendry}\ \emph {et~al.}(2015)\citenamefont {Pendry},
  \citenamefont {Luo},\ and\ \citenamefont {Zhao}}]{Pendry:15}%
  \BibitemOpen
  \bibfield  {author} {\bibinfo {author} {\bibfnamefont {J.~B.}\ \bibnamefont
  {Pendry}}, \bibinfo {author} {\bibfnamefont {Y.}~\bibnamefont {Luo}}, \ and\
  \bibinfo {author} {\bibfnamefont {R.}~\bibnamefont {Zhao}},\ }\href@noop {}
  {\bibfield  {journal} {\bibinfo  {journal} {Science}\ }\textbf {\bibinfo
  {volume} {348}},\ \bibinfo {pages} {521} (\bibinfo {year}
  {2015})}\BibitemShut {NoStop}%
\bibitem [{\citenamefont {Aubry}\ \emph
  {et~al.}(2010{\natexlab{b}})\citenamefont {Aubry}, \citenamefont {Lei},
  \citenamefont {Maier},\ and\ \citenamefont {Pendry}}]{Aubry:10}%
  \BibitemOpen
  \bibfield  {author} {\bibinfo {author} {\bibfnamefont {A.}~\bibnamefont
  {Aubry}}, \bibinfo {author} {\bibfnamefont {D.~Y.}\ \bibnamefont {Lei}},
  \bibinfo {author} {\bibfnamefont {S.~A.}\ \bibnamefont {Maier}}, \ and\
  \bibinfo {author} {\bibfnamefont {J.~B.}\ \bibnamefont {Pendry}},\
  }\href@noop {} {\bibfield  {journal} {\bibinfo  {journal} {Phys. Rev. B}\
  }\textbf {\bibinfo {volume} {82}},\ \bibinfo {pages} {205109} (\bibinfo
  {year} {2010}{\natexlab{b}})}\BibitemShut {NoStop}%
\bibitem [{\citenamefont {Fern{\'a}ndez-Dom{\'\i}nguez}\ \emph
  {et~al.}(2012)\citenamefont {Fern{\'a}ndez-Dom{\'\i}nguez}, \citenamefont
  {Zhang}, \citenamefont {Luo}, \citenamefont {Maier}, \citenamefont
  {Garc{\'\i}a-Vidal},\ and\ \citenamefont {Pendry}}]{Fernandez:12}%
  \BibitemOpen
  \bibfield  {author} {\bibinfo {author} {\bibfnamefont {A.~I.}\ \bibnamefont
  {Fern{\'a}ndez-Dom{\'\i}nguez}}, \bibinfo {author} {\bibfnamefont
  {P.}~\bibnamefont {Zhang}}, \bibinfo {author} {\bibfnamefont
  {Y.}~\bibnamefont {Luo}}, \bibinfo {author} {\bibfnamefont {S.~A.}\
  \bibnamefont {Maier}}, \bibinfo {author} {\bibfnamefont {F.~J.}\ \bibnamefont
  {Garc{\'\i}a-Vidal}}, \ and\ \bibinfo {author} {\bibfnamefont {J.~B.}\
  \bibnamefont {Pendry}},\ }\href@noop {} {\bibfield  {journal} {\bibinfo
  {journal} {Phys. Rev. B}\ }\textbf {\bibinfo {volume} {86}},\ \bibinfo
  {pages} {241110} (\bibinfo {year} {2012})}\BibitemShut {NoStop}%
\bibitem [{\citenamefont {Pendry}\ \emph {et~al.}(2013)\citenamefont {Pendry},
  \citenamefont {Fern{\'a}ndez-Dom{\'\i}nguez}, \citenamefont {Luo},\ and\
  \citenamefont {Zhao}}]{Pendry:13}%
  \BibitemOpen
  \bibfield  {author} {\bibinfo {author} {\bibfnamefont {J.~B.}\ \bibnamefont
  {Pendry}}, \bibinfo {author} {\bibfnamefont {A.~I.}\ \bibnamefont
  {Fern{\'a}ndez-Dom{\'\i}nguez}}, \bibinfo {author} {\bibfnamefont
  {Y.}~\bibnamefont {Luo}}, \ and\ \bibinfo {author} {\bibfnamefont
  {R.}~\bibnamefont {Zhao}},\ }\href@noop {} {\bibfield  {journal} {\bibinfo
  {journal} {Nat. Phys.}\ }\textbf {\bibinfo {volume} {9}},\ \bibinfo {pages}
  {518} (\bibinfo {year} {2013})}\BibitemShut {NoStop}%
\bibitem [{\citenamefont {Lebedev}\ \emph {et~al.}(2010)\citenamefont
  {Lebedev}, \citenamefont {Vergeles},\ and\ \citenamefont
  {Vorobev}}]{Lebedev:10}%
  \BibitemOpen
  \bibfield  {author} {\bibinfo {author} {\bibfnamefont {V.}~\bibnamefont
  {Lebedev}}, \bibinfo {author} {\bibfnamefont {S.}~\bibnamefont {Vergeles}}, \
  and\ \bibinfo {author} {\bibfnamefont {P.}~\bibnamefont {Vorobev}},\
  }\href@noop {} {\bibfield  {journal} {\bibinfo  {journal} {Opt. Lett.}\
  }\textbf {\bibinfo {volume} {35}},\ \bibinfo {pages} {640} (\bibinfo {year}
  {2010})}\BibitemShut {NoStop}%
\bibitem [{\citenamefont {Klimov}\ and\ \citenamefont
  {Guzatov}(2007)}]{Klimov:07}%
  \BibitemOpen
  \bibfield  {author} {\bibinfo {author} {\bibfnamefont {V.~V.}\ \bibnamefont
  {Klimov}}\ and\ \bibinfo {author} {\bibfnamefont {D.~V.}\ \bibnamefont
  {Guzatov}},\ }\href@noop {} {\bibfield  {journal} {\bibinfo  {journal} {Phys.
  Rev. B}\ }\textbf {\bibinfo {volume} {75}},\ \bibinfo {pages} {024303}
  (\bibinfo {year} {2007})}\BibitemShut {NoStop}%
\bibitem [{\citenamefont {Vorobev}(2010)}]{Vorobev:10}%
  \BibitemOpen
  \bibfield  {author} {\bibinfo {author} {\bibfnamefont {P.~E.}\ \bibnamefont
  {Vorobev}},\ }\href@noop {} {\bibfield  {journal} {\bibinfo  {journal} {J.
  Exp. Theor. Phys.}\ }\textbf {\bibinfo {volume} {110}},\ \bibinfo {pages}
  {193} (\bibinfo {year} {2010})}\BibitemShut {NoStop}%
\bibitem [{\citenamefont {Lebedev}\ \emph {et~al.}(2013)\citenamefont
  {Lebedev}, \citenamefont {Vergeles},\ and\ \citenamefont
  {Vorobev}}]{Lebedev:13}%
  \BibitemOpen
  \bibfield  {author} {\bibinfo {author} {\bibfnamefont {V.~V.}\ \bibnamefont
  {Lebedev}}, \bibinfo {author} {\bibfnamefont {S.~S.}\ \bibnamefont
  {Vergeles}}, \ and\ \bibinfo {author} {\bibfnamefont {P.~E.}\ \bibnamefont
  {Vorobev}},\ }\href@noop {} {\bibfield  {journal} {\bibinfo  {journal} {Appl.
  Phys. B}\ }\textbf {\bibinfo {volume} {111}},\ \bibinfo {pages} {577}
  (\bibinfo {year} {2013})}\BibitemShut {NoStop}%
\bibitem [{\citenamefont {Hinch}(1991)}]{Hinch:91}%
  \BibitemOpen
  \bibfield  {author} {\bibinfo {author} {\bibfnamefont {E.~J.}\ \bibnamefont
  {Hinch}},\ }\href@noop {} {\emph {\bibinfo {title} {Perturbation methods}}}\
  (\bibinfo  {publisher} {Cambridge university press},\ \bibinfo {year}
  {1991})\BibitemShut {NoStop}%
\bibitem [{\citenamefont {Ruppin}(1989)}]{Ruppin:89}%
  \BibitemOpen
  \bibfield  {author} {\bibinfo {author} {\bibfnamefont {R.}~\bibnamefont
  {Ruppin}},\ }\href@noop {} {\bibfield  {journal} {\bibinfo  {journal} {J.
  Phys. Soc. Jpn.}\ }\textbf {\bibinfo {volume} {58}},\ \bibinfo {pages} {1446}
  (\bibinfo {year} {1989})}\BibitemShut {NoStop}%
\bibitem [{\citenamefont {Tamaru}\ \emph {et~al.}(2002)\citenamefont {Tamaru},
  \citenamefont {Kuwata}, \citenamefont {Miyazaki},\ and\ \citenamefont
  {Miyano}}]{Tamaru:02}%
  \BibitemOpen
  \bibfield  {author} {\bibinfo {author} {\bibfnamefont {H.}~\bibnamefont
  {Tamaru}}, \bibinfo {author} {\bibfnamefont {H.}~\bibnamefont {Kuwata}},
  \bibinfo {author} {\bibfnamefont {H.~T.}\ \bibnamefont {Miyazaki}}, \ and\
  \bibinfo {author} {\bibfnamefont {K.}~\bibnamefont {Miyano}},\ }\href@noop {}
  {\bibfield  {journal} {\bibinfo  {journal} {Appl. Phys. Lett.}\ }\textbf
  {\bibinfo {volume} {80}},\ \bibinfo {pages} {1826} (\bibinfo {year}
  {2002})}\BibitemShut {NoStop}%
\bibitem [{\citenamefont {Rechberger}\ \emph {et~al.}(2003)\citenamefont
  {Rechberger}, \citenamefont {Hohenau}, \citenamefont {Leitner}, \citenamefont
  {Krenn}, \citenamefont {Lamprecht},\ and\ \citenamefont
  {Aussenegg}}]{Rechberger:03}%
  \BibitemOpen
  \bibfield  {author} {\bibinfo {author} {\bibfnamefont {W.}~\bibnamefont
  {Rechberger}}, \bibinfo {author} {\bibfnamefont {A.}~\bibnamefont {Hohenau}},
  \bibinfo {author} {\bibfnamefont {A.}~\bibnamefont {Leitner}}, \bibinfo
  {author} {\bibfnamefont {J.~R.}\ \bibnamefont {Krenn}}, \bibinfo {author}
  {\bibfnamefont {B.}~\bibnamefont {Lamprecht}}, \ and\ \bibinfo {author}
  {\bibfnamefont {F.~R.}\ \bibnamefont {Aussenegg}},\ }\href@noop {} {\bibfield
   {journal} {\bibinfo  {journal} {Opt Commun}\ }\textbf {\bibinfo {volume}
  {220}},\ \bibinfo {pages} {137} (\bibinfo {year} {2003})}\BibitemShut
  {NoStop}%
\bibitem [{\citenamefont {Gunnarsson}\ \emph {et~al.}(2005)\citenamefont
  {Gunnarsson}, \citenamefont {Rindzevicius}, \citenamefont {Prikulis},
  \citenamefont {Kasemo}, \citenamefont {K{\"a}ll}, \citenamefont {Zou},\ and\
  \citenamefont {Schatz}}]{Gunnarsson:05}%
  \BibitemOpen
  \bibfield  {author} {\bibinfo {author} {\bibfnamefont {L.}~\bibnamefont
  {Gunnarsson}}, \bibinfo {author} {\bibfnamefont {T.}~\bibnamefont
  {Rindzevicius}}, \bibinfo {author} {\bibfnamefont {J.}~\bibnamefont
  {Prikulis}}, \bibinfo {author} {\bibfnamefont {B.}~\bibnamefont {Kasemo}},
  \bibinfo {author} {\bibfnamefont {M.}~\bibnamefont {K{\"a}ll}}, \bibinfo
  {author} {\bibfnamefont {S.}~\bibnamefont {Zou}}, \ and\ \bibinfo {author}
  {\bibfnamefont {G.~C.}\ \bibnamefont {Schatz}},\ }\href@noop {} {\bibfield
  {journal} {\bibinfo  {journal} {J. Phys. Chem. B}\ }\textbf {\bibinfo
  {volume} {109}},\ \bibinfo {pages} {1079} (\bibinfo {year}
  {2005})}\BibitemShut {NoStop}%
\bibitem [{\citenamefont {Jain}\ \emph {et~al.}(2007)\citenamefont {Jain},
  \citenamefont {Huang},\ and\ \citenamefont {El-Sayed}}]{Jain:07}%
  \BibitemOpen
  \bibfield  {author} {\bibinfo {author} {\bibfnamefont {P.~K.}\ \bibnamefont
  {Jain}}, \bibinfo {author} {\bibfnamefont {W.}~\bibnamefont {Huang}}, \ and\
  \bibinfo {author} {\bibfnamefont {M.~A.}\ \bibnamefont {El-Sayed}},\
  }\href@noop {} {\bibfield  {journal} {\bibinfo  {journal} {Nano Lett.}\
  }\textbf {\bibinfo {volume} {7}},\ \bibinfo {pages} {2080} (\bibinfo {year}
  {2007})}\BibitemShut {NoStop}%
\bibitem [{\citenamefont {Kadkhodazadeh}\ \emph {et~al.}(2014)\citenamefont
  {Kadkhodazadeh}, \citenamefont {de~Lasson}, \citenamefont {Beleggia},
  \citenamefont {Kneipp}, \citenamefont {Wagner},\ and\ \citenamefont
  {Kneipp}}]{Kadkhodazadeh:14}%
  \BibitemOpen
  \bibfield  {author} {\bibinfo {author} {\bibfnamefont {S.}~\bibnamefont
  {Kadkhodazadeh}}, \bibinfo {author} {\bibfnamefont {J.~R.}\ \bibnamefont
  {de~Lasson}}, \bibinfo {author} {\bibfnamefont {M.}~\bibnamefont {Beleggia}},
  \bibinfo {author} {\bibfnamefont {H.}~\bibnamefont {Kneipp}}, \bibinfo
  {author} {\bibfnamefont {J.~B.}\ \bibnamefont {Wagner}}, \ and\ \bibinfo
  {author} {\bibfnamefont {K.}~\bibnamefont {Kneipp}},\ }\href@noop {}
  {\bibfield  {journal} {\bibinfo  {journal} {J. Phys. Chem. C}\ }\textbf
  {\bibinfo {volume} {118}},\ \bibinfo {pages} {5478} (\bibinfo {year}
  {2014})}\BibitemShut {NoStop}%
\bibitem [{\citenamefont {Landau}\ \emph {et~al.}(1984)\citenamefont {Landau},
  \citenamefont {Bell}, \citenamefont {Kearsley}, \citenamefont {Pitaevskii},
  \citenamefont {Lifshitz},\ and\ \citenamefont {Sykes}}]{Landau:Book}%
  \BibitemOpen
  \bibfield  {author} {\bibinfo {author} {\bibfnamefont {L.~D.}\ \bibnamefont
  {Landau}}, \bibinfo {author} {\bibfnamefont {J.~S.}\ \bibnamefont {Bell}},
  \bibinfo {author} {\bibfnamefont {M.~J.}\ \bibnamefont {Kearsley}}, \bibinfo
  {author} {\bibfnamefont {L.~P.}\ \bibnamefont {Pitaevskii}}, \bibinfo
  {author} {\bibfnamefont {E.~M.}\ \bibnamefont {Lifshitz}}, \ and\ \bibinfo
  {author} {\bibfnamefont {J.~B.}\ \bibnamefont {Sykes}},\ }\href@noop {}
  {\emph {\bibinfo {title} {Electrodynamics of continuous media}}},\
  Vol.~\bibinfo {volume} {8}\ (\bibinfo  {publisher} {elsevier},\ \bibinfo
  {year} {1984})\BibitemShut {NoStop}%
\bibitem [{\citenamefont {Ando}\ and\ \citenamefont {Kang}(2014)}]{Ando:14}%
  \BibitemOpen
  \bibfield  {author} {\bibinfo {author} {\bibfnamefont {K.}~\bibnamefont
  {Ando}}\ and\ \bibinfo {author} {\bibfnamefont {H.}~\bibnamefont {Kang}},\
  }\href@noop {} {\bibfield  {journal} {\bibinfo  {journal} {arXiv preprint
  arXiv:1412.6250}\ } (\bibinfo {year} {2014})}\BibitemShut {NoStop}%
\bibitem [{\citenamefont {Grieser}(2014)}]{Grieser:14}%
  \BibitemOpen
  \bibfield  {author} {\bibinfo {author} {\bibfnamefont {D.}~\bibnamefont
  {Grieser}},\ }\href@noop {} {\bibfield  {journal} {\bibinfo  {journal} {Rev.
  Math. Phys.}\ }\textbf {\bibinfo {volume} {26}},\ \bibinfo {pages} {1450005}
  (\bibinfo {year} {2014})}\BibitemShut {NoStop}%
\bibitem [{\citenamefont {Ammari}\ \emph {et~al.}(2015)\citenamefont {Ammari},
  \citenamefont {Millien}, \citenamefont {Ruiz},\ and\ \citenamefont
  {Zhang}}]{Ammari:15}%
  \BibitemOpen
  \bibfield  {author} {\bibinfo {author} {\bibfnamefont {H.}~\bibnamefont
  {Ammari}}, \bibinfo {author} {\bibfnamefont {P.}~\bibnamefont {Millien}},
  \bibinfo {author} {\bibfnamefont {M.}~\bibnamefont {Ruiz}}, \ and\ \bibinfo
  {author} {\bibfnamefont {H.}~\bibnamefont {Zhang}},\ }\href@noop {}
  {\bibfield  {journal} {\bibinfo  {journal} {arXiv preprint arXiv:1506.00866}\
  } (\bibinfo {year} {2015})}\BibitemShut {NoStop}%
\bibitem [{\citenamefont {Jeffrey}\ and\ \citenamefont
  {Van~Dyke}(1978)}]{Jeffrey:78}%
  \BibitemOpen
  \bibfield  {author} {\bibinfo {author} {\bibfnamefont {D.~J.}\ \bibnamefont
  {Jeffrey}}\ and\ \bibinfo {author} {\bibfnamefont {M.}~\bibnamefont
  {Van~Dyke}},\ }\href@noop {} {\bibfield  {journal} {\bibinfo  {journal} {IMA
  J. Appl. Math.}\ }\textbf {\bibinfo {volume} {22}},\ \bibinfo {pages} {337}
  (\bibinfo {year} {1978})}\BibitemShut {NoStop}%
\bibitem [{\citenamefont {Moon}\ and\ \citenamefont {Spencer}(1961)}]{Moon:61}%
  \BibitemOpen
  \bibfield  {author} {\bibinfo {author} {\bibfnamefont {P.~H.}\ \bibnamefont
  {Moon}}\ and\ \bibinfo {author} {\bibfnamefont {D.~E.}\ \bibnamefont
  {Spencer}},\ }\href@noop {} {\emph {\bibinfo {title} {Field Theory Handbook:
  Including Coordinate Systems, Differential Equations, and Their Solutions}}}\
  (\bibinfo  {publisher} {Springer-Verlag},\ \bibinfo {year}
  {1961})\BibitemShut {NoStop}%
\bibitem [{\citenamefont {Sneddon}(1972)}]{Sneddon:Book}%
  \BibitemOpen
  \bibfield  {author} {\bibinfo {author} {\bibfnamefont {I.~N.}\ \bibnamefont
  {Sneddon}},\ }\href@noop {} {\emph {\bibinfo {title} {The use of integral
  transforms}}}\ (\bibinfo  {publisher} {McGraw-Hill},\ \bibinfo {year}
  {1972})\BibitemShut {NoStop}%
\bibitem [{\citenamefont {Crighton}\ \emph {et~al.}(2012)\citenamefont
  {Crighton}, \citenamefont {Dowling}, \citenamefont {Williams}, \citenamefont
  {Heckl},\ and\ \citenamefont {Leppington}}]{Crighton:12}%
  \BibitemOpen
  \bibfield  {author} {\bibinfo {author} {\bibfnamefont {D.~G.}\ \bibnamefont
  {Crighton}}, \bibinfo {author} {\bibfnamefont {A.~P.}\ \bibnamefont
  {Dowling}}, \bibinfo {author} {\bibfnamefont {J.~E.~F.}\ \bibnamefont
  {Williams}}, \bibinfo {author} {\bibfnamefont {M.~A.}\ \bibnamefont {Heckl}},
  \ and\ \bibinfo {author} {\bibfnamefont {F.~A.}\ \bibnamefont {Leppington}},\
  }\href@noop {} {\emph {\bibinfo {title} {Modern methods in analytical
  acoustics: lecture notes}}}\ (\bibinfo  {publisher} {Springer Science \&
  Business Media},\ \bibinfo {year} {2012})\BibitemShut {NoStop}%
\bibitem [{\citenamefont {Abramowitz}\ and\ \citenamefont
  {Stegun}()}]{Abramowitz:book}%
  \BibitemOpen
  \bibfield  {author} {\bibinfo {author} {\bibfnamefont {M.}~\bibnamefont
  {Abramowitz}}\ and\ \bibinfo {author} {\bibfnamefont {I.~A.}\ \bibnamefont
  {Stegun}},\ }\href@noop {} {\emph {\bibinfo {title} {Handbook of mathematical
  functions}}},\ Vol.~\bibinfo {volume} {1}\BibitemShut {NoStop}%
\bibitem [{Note1()}]{Note1}%
  \BibitemOpen
  \bibinfo {note} {Eq.~\protect \textup {\hbox {\mathsurround \z@ \protect
  \normalfont (\ignorespaces \ref {trans}\unskip \@@italiccorr )}} produces one
  additional negative solution. The latter branch, however, approaches $-\infty
  $ as $h\to 0$, whereby \protect \textup {\hbox {\mathsurround \z@ \protect
  \normalfont (\ignorespaces \ref {algebra result}\unskip \@@italiccorr )}}
  shows that $\lambda $ becomes small and asymptoticness is lost. Consistency
  requires disregarding this spurious mode, in agreement with general
  considerations}\BibitemShut {NoStop}%
\bibitem [{\citenamefont {Johnson}\ and\ \citenamefont
  {Christy}(1972)}]{Johnson:72}%
  \BibitemOpen
  \bibfield  {author} {\bibinfo {author} {\bibfnamefont {P.~B.}\ \bibnamefont
  {Johnson}}\ and\ \bibinfo {author} {\bibfnamefont {R.~W.}\ \bibnamefont
  {Christy}},\ }\href@noop {} {\bibfield  {journal} {\bibinfo  {journal} {Phys.
  Rev. B}\ }\textbf {\bibinfo {volume} {6}},\ \bibinfo {pages} {4370} (\bibinfo
  {year} {1972})}\BibitemShut {NoStop}%
\bibitem [{\citenamefont {Goyette}\ and\ \citenamefont
  {Navon}(1976)}]{Goyette:76}%
  \BibitemOpen
  \bibfield  {author} {\bibinfo {author} {\bibfnamefont {A.}~\bibnamefont
  {Goyette}}\ and\ \bibinfo {author} {\bibfnamefont {A.}~\bibnamefont
  {Navon}},\ }\href@noop {} {\bibfield  {journal} {\bibinfo  {journal} {Phys.
  Rev. B}\ }\textbf {\bibinfo {volume} {13}},\ \bibinfo {pages} {4320}
  (\bibinfo {year} {1976})}\BibitemShut {NoStop}%
\bibitem [{\citenamefont {Klimov}\ and\ \citenamefont
  {Lambrecht}(2009)}]{Klimov:09}%
  \BibitemOpen
  \bibfield  {author} {\bibinfo {author} {\bibfnamefont {V.~V.}\ \bibnamefont
  {Klimov}}\ and\ \bibinfo {author} {\bibfnamefont {A.}~\bibnamefont
  {Lambrecht}},\ }\href@noop {} {\bibfield  {journal} {\bibinfo  {journal}
  {Plasmonics}\ }\textbf {\bibinfo {volume} {4}},\ \bibinfo {pages} {31}
  (\bibinfo {year} {2009})}\BibitemShut {NoStop}%
\bibitem [{\citenamefont {R.}\ \emph {et~al.}(2013)\citenamefont {R.},
  \citenamefont {C.}, \citenamefont {Somerville},\ and\ \citenamefont
  {Augui{\'e}}}]{Le:13}%
  \BibitemOpen
  \bibfield  {author} {\bibinfo {author} {\bibfnamefont {L.}~\bibnamefont
  {R.}}, \bibinfo {author} {\bibfnamefont {E.}~\bibnamefont {C.}}, \bibinfo
  {author} {\bibfnamefont {W.~R.~C.}\ \bibnamefont {Somerville}}, \ and\
  \bibinfo {author} {\bibfnamefont {B.}~\bibnamefont {Augui{\'e}}},\
  }\href@noop {} {\bibfield  {journal} {\bibinfo  {journal} {Phys. Rev. A}\
  }\textbf {\bibinfo {volume} {87}},\ \bibinfo {pages} {012504} (\bibinfo
  {year} {2013})}\BibitemShut {NoStop}%
\bibitem [{\citenamefont {Ando}\ \emph {et~al.}(2015)\citenamefont {Ando},
  \citenamefont {Kang},\ and\ \citenamefont {Liu}}]{Ando:15}%
  \BibitemOpen
  \bibfield  {author} {\bibinfo {author} {\bibfnamefont {K.}~\bibnamefont
  {Ando}}, \bibinfo {author} {\bibfnamefont {H.}~\bibnamefont {Kang}}, \ and\
  \bibinfo {author} {\bibfnamefont {H.}~\bibnamefont {Liu}},\ }\href@noop {}
  {\bibfield  {journal} {\bibinfo  {journal} {arXiv preprint arXiv:1506.03566}\
  } (\bibinfo {year} {2015})}\BibitemShut {NoStop}%
\bibitem [{\citenamefont {Luo}\ \emph {et~al.}(2013)\citenamefont {Luo},
  \citenamefont {Fernandez-Dominguez}, \citenamefont {Wiener}, \citenamefont
  {Maier},\ and\ \citenamefont {Pendry}}]{Luo:13}%
  \BibitemOpen
  \bibfield  {author} {\bibinfo {author} {\bibfnamefont {Y.}~\bibnamefont
  {Luo}}, \bibinfo {author} {\bibfnamefont {A.}~\bibnamefont
  {Fernandez-Dominguez}}, \bibinfo {author} {\bibfnamefont {A.}~\bibnamefont
  {Wiener}}, \bibinfo {author} {\bibfnamefont {S.~A.}\ \bibnamefont {Maier}}, \
  and\ \bibinfo {author} {\bibfnamefont {J.~B.}\ \bibnamefont {Pendry}},\
  }\href@noop {} {\bibfield  {journal} {\bibinfo  {journal} {Phys. Rev. Lett.}\
  }\textbf {\bibinfo {volume} {111}},\ \bibinfo {pages} {093901} (\bibinfo
  {year} {2013})}\BibitemShut {NoStop}%
\bibitem [{\citenamefont {Cirac{\`\i}}\ \emph {et~al.}(2013)\citenamefont
  {Cirac{\`\i}}, \citenamefont {Pendry},\ and\ \citenamefont
  {Smith}}]{Ciraci:13}%
  \BibitemOpen
  \bibfield  {author} {\bibinfo {author} {\bibfnamefont {C.}~\bibnamefont
  {Cirac{\`\i}}}, \bibinfo {author} {\bibfnamefont {J.~B.}\ \bibnamefont
  {Pendry}}, \ and\ \bibinfo {author} {\bibfnamefont {D.~R.}\ \bibnamefont
  {Smith}},\ }\href@noop {} {\bibfield  {journal} {\bibinfo  {journal}
  {ChemPhysChem}\ }\textbf {\bibinfo {volume} {14}},\ \bibinfo {pages} {1109}
  (\bibinfo {year} {2013})}\BibitemShut {NoStop}%
\bibitem [{\citenamefont {Raza}\ \emph {et~al.}(2015)\citenamefont {Raza},
  \citenamefont {Bozhevolnyi}, \citenamefont {Wubs},\ and\ \citenamefont
  {Mortensen}}]{Raza:15}%
  \BibitemOpen
  \bibfield  {author} {\bibinfo {author} {\bibfnamefont {S.}~\bibnamefont
  {Raza}}, \bibinfo {author} {\bibfnamefont {S.~I.}\ \bibnamefont
  {Bozhevolnyi}}, \bibinfo {author} {\bibfnamefont {M.}~\bibnamefont {Wubs}}, \
  and\ \bibinfo {author} {\bibfnamefont {N.~A.}\ \bibnamefont {Mortensen}},\
  }\href@noop {} {\bibfield  {journal} {\bibinfo  {journal} {J. Phys. Condens.
  Matter}\ }\textbf {\bibinfo {volume} {27}},\ \bibinfo {pages} {183204}
  (\bibinfo {year} {2015})}\BibitemShut {NoStop}%
\bibitem [{\citenamefont {Schnitzer}\ \emph {et~al.}(2015)\citenamefont
  {Schnitzer}, \citenamefont {Giannini}, \citenamefont {Craster},\ and\
  \citenamefont {Maier}}]{Schnitzer:15arxivB}%
  \BibitemOpen
  \bibfield  {author} {\bibinfo {author} {\bibfnamefont {O.}~\bibnamefont
  {Schnitzer}}, \bibinfo {author} {\bibfnamefont {V.}~\bibnamefont {Giannini}},
  \bibinfo {author} {\bibfnamefont {R.~V.}\ \bibnamefont {Craster}}, \ and\
  \bibinfo {author} {\bibfnamefont {S.~A.}\ \bibnamefont {Maier}},\ }\href@noop
  {} {\bibfield  {journal} {\bibinfo  {journal} {arXiv preprint
  arXiv:1511.04895}\ } (\bibinfo {year} {2015})}\BibitemShut {NoStop}%
\bibitem [{\citenamefont {Wolfram~Research}(2015)}]{Wolfram}%
  \BibitemOpen
  \bibfield  {author} {\bibinfo {author} {\bibfnamefont {I.}~\bibnamefont
  {Wolfram~Research}},\ }\href@noop {} {\enquote {\bibinfo {title} {Mathematica
  10.1},}\ } (\bibinfo {year} {2015})\BibitemShut {NoStop}%
\bibitem [{\citenamefont {Olver}(2010)}]{Olver:Book}%
  \BibitemOpen
  \bibfield  {author} {\bibinfo {author} {\bibfnamefont {F.~W.~J.}\
  \bibnamefont {Olver}},\ }\href@noop {} {\emph {\bibinfo {title} {NIST
  handbook of mathematical functions}}}\ (\bibinfo  {publisher} {Cambridge
  University Press},\ \bibinfo {year} {2010})\BibitemShut {NoStop}%
\end{thebibliography}%
\end{document}